\documentclass{article}

\usepackage{arXiv}

\usepackage[utf8]{inputenc} 
\usepackage[T1]{fontenc}    

\usepackage{xcolor}
\usepackage[colorlinks = true,
            linkcolor = gray,
            urlcolor  = gray,
            citecolor = gray,
            anchorcolor = red]{hyperref}
            
\usepackage{url}            

\usepackage{booktabs}       
\usepackage{multirow}       
\usepackage{tabularx}       
\usepackage{makecell}       
\newcolumntype{C}{>{\centering\arraybackslash}X} 

\usepackage{amsmath}        
\usepackage{amsfonts}       
\usepackage{nicefrac}       

\usepackage{microtype}      

\usepackage{apacite} 
\usepackage{doi}            

\usepackage{graphicx}       
\usepackage{float}          

\usepackage{fancyhdr}       
\title{
Improving AI weather prediction models using global mass and energy conservation schemes
    \thanks{Sha, Y. et al. Improving AI weather prediction models using global mass and energy conservation schemes. Submitted for publication in Journal of Advances in Modeling Earth Systems (JAMES)}
    \author{
      \textbf{Yingkai Sha}$^{\dagger}$,
      \textbf{John S. Schreck}$^{\dagger}$,
      \textbf{William Chapman}$^{\ddagger}$,
      \textbf{David John Gagne II}$^{\dagger}$\\[1em]
      Computational and Information Systems (CISL) Laboratory$^{\dagger}$ \\
      Climate and Global Dynamics (CGD) Laboratory$^{\ddagger}$ \\[1em]
      NSF National Center for Atmospheric Research \\
      Boulder, Colorado, USA\\[1em]
      \texttt{\{ksha, schreck, wchapman, dgagne\}@ucar.edu}
    }
}

\begin{document}
\maketitle

\begin{abstract}
Artificial Intelligence (AI) weather prediction (AIWP) models are powerful tools for medium-range forecasts but often lack physical consistency, leading to outputs that violate conservation laws. This study introduces a set of novel physics-based schemes designed to enforce the conservation of global dry air mass, moisture budget, and total atmospheric energy in AIWP models. The schemes are highly modular, allowing for seamless integration into a wide range of AI model architectures. Forecast experiments are conducted to demonstrate the benefit of conservation schemes using FuXi, an example AIWP model, modified and adapted for 1.0$^\circ$ grid spacing. Verification results show that the conservation schemes can guide the model in producing forecasts that obey conservation laws. The forecast skills of upper-air and surface variables are also improved, with longer forecast lead times receiving larger benefits. Notably, large performance gains are found in the total precipitation forecasts, owing to the reduction of drizzle bias. The proposed conservation schemes establish a foundation for implementing other physics-based schemes in the future. They also provide a new way to integrate atmospheric domain knowledge into the design and refinement of AIWP models.
\end{abstract}

\section*{Plain Language Summary}
Artificial intelligence (AI) models can learn how to make weather forecasts directly from historical analysis data. They are known as AI weather prediction models (AIWP). These models are purely data-driven, and their forecasts do not obey global mass and energy conservations. This study provides numerical schemes to guide AIWP models in producing forecasts that obey conservation laws. We compared the results from two AIWP model runs; one is purely data-driven, and the other has our numerical schemes. The latter conserves mass and energy better; their forecast skills are also improved. This shows that our numerical schemes are successful. This also means that AIWP models can potentially receive more benefits from other numerical schemes in the future.


%
%

\section{Introduction}

Artificial Intelligence (AI) weather prediction (hereafter AIWP) models have attracted widespread attention from the atmospheric science community following the successful releases of Pangu Weather \cite{bi2023accurate}, GraphCast \cite{lam2023learning}, and many others \cite<e.g.>{chen2023fuxi,nguyen2023climax,bonev2023spherical,bodnar2024aurora,nguyen2023scaling,willard2024analyzing,lang2024aifs,schreck2024community}. By leveraging Transformer-based and graph-based neural network architectures and high-quality datasets, such as the European Centre for Medium-Range Weather Forecasts (ECMWF) Reanalysis version 5 \cite<ERA5; >{hersbach2020era5}, these models have demonstrated competitive forecast performance. In many cases, they outperform state-of-the-art global Numerical Weather Prediction (NWP) models in deterministic verification against individual prognostic variables. Despite their initial successes, users have found limitations in existing AIWP models that complicate their ability to be fully integrated into the traditional NWP pipeline. For example, AIWP models can produce unphysical outputs, such as negative humidity \cite<e.g.>{schreck2024community}, they have stability problems for longer simulations \cite{watt2023ace,slivinski2024assimilating}, and they may lack fidelity on certain process-based verification aspects [e.g., \citeA{selz2023can} for the butterfly effect; \citeA{bonavita2024some} for the geostrophic balance]. Many of these limitations are rooted in the lack of consideration of physical relationships in AI models.

Efforts to address the absence of physical relationships in AI models have historically centered on two areas: Physics-Informed Neural Networks (PINNs) and hybrid models. PINNs enforce physical relationships expressed as Partial Differential Equations (PDEs) by either incorporating PDE-defined loss terms during training [e.g., \citeA{raissi2019physics,torres2022mesh,jagtap2020conservative} for using collocation points] or embedding PDE relationships directly into the model design [e.g., \citeA{richter2022neural} for using stream function to model incompressible fluid]. Hybrid models combine numerical dynamic cores with neural-network-based subgrid-scale parameterizations, where the latter are constrained to satisfy physical properties [e.g., \citeA{beucler2021enforcing,yu2024climsim} for the conservation of mass and energy; \citeA{hu2024stable} for liquid and ice cloud partitions]. 

The application of physical constraints in AI weather and climate models is in its early stages. Initial progress has been made by the AI2 Climate Emulator \cite{watt2023ace,watt2024ace2}, primarily focused on simulations on decadal time scales. For medium-range weather forecasts, most AIWP models do not have any physical constraints. Several studies have pointed to the demand for such a move. \citeA{bi2023accurate} discussed the need for ``real-world constraints'' to improve Pangu Weather on longer forecast lead times. NeuralGCM \cite{kochkov2024neural} has also created space for physics integrations in the future using the hybrid modeling approach and has recently incorporated satellite-derived precipitation \cite{yuval2024neural}. However, the full benefit and potential impact of incorporating physical constraints in AIWP models remains underexplored.

To meet the critical need of applying physical constraints in AIWP models, this study introduces a set of novel physics-based schemes specifically designed for medium-range AIWP models. These schemes perform residual corrections on the global dry air mass content, moisture budgets, and total atmospheric energy in AIWP models and guide them to produce forecasts that better reflect conservation laws. The main challenge of developing these physics-based schemes is the technical gap between the formulation of existing AIWP models and the physical constraints established in numerical systems. AIWP models are typically developed on constant pressure levels; they have limited output variables to describe the state of the atmosphere, and their internal calculations are purely data-driven--- these characteristics differ highly from NWP models, which in many ways have been designed to respect physical constraints. To tackle this challenge, our proposed conservation schemes are highly modular, making them adaptable to different combinations of AI architectures, spatial resolutions, and training workflows. To demonstrate their effectiveness, we select FuXi, an AIWP model benchmarked on \citeA{rasp2024weatherbench}, as an example case. The ERA5, re-gridded to 1.0$^\circ$ grid spacing, serves as the training and verification dataset. The implementation of the research ideas above is based on the Community Research Earth Digital Intelligence Twin \cite<CREDIT; >{schreck2024community} platform hosted at NSF National Center for Atmospheric Research (NCAR).

This study addresses the following scientific questions: (1) Can conservation schemes improve the forecast skill and conservation behaviors of AIWP models? (2) Which AIWP output variables benefit the most from conservation schemes? (3) What is the relative importance of conservation laws in medium-range AIWP? By answering these questions, this study will present an effective solution for how physical constraints, such as global mass and energy conservations, can be adapted into AIWP models. Through numerical experiments and verifications, this study will also show how physical constraints can benefit AIWP models and guide them in producing better forecasts. We also hope this study raises awareness of the importance of atmospheric science domain knowledge in AI-based weather forecasting and inspires future innovations in developing AIWP models that better respect physical relationships. 


\section{Data}\label{sec2}

\subsection{The ERA5 dataset}\label{sec21}

\begin{table}
\begin{center}
\caption{The variables of interest in this study.}\label{tab1}
\renewcommand{\arraystretch}{1.2}
\begin{tabularx}{\textwidth}
{c >{\centering\arraybackslash}X c c}
\specialrule{1.5pt}{0pt}{3pt}
Type & Variable Name & Units & Role \\ 
\midrule
\multirow{5}{*}{Pressure level}
& Zonal Wind                                 & $\mathrm{m \cdot s^{-1}}$   & \multirow{5}{*}{Prognostic, Instantaneous} \\
& Meridional Wind                            & $\mathrm{m \cdot s^{-1}}$   &\\
& Air Temperature                            & $\mathrm{K}$                &\\
& Specific Total Water\textsuperscript{a}    & $\mathrm{kg \cdot kg^{-1}}$ &\\
& Geopotential height                        & $\mathrm{m}$                &\\
\midrule
\multirow{4}{*}{Single level}& Mean Sea Level Pressure                    & $\mathrm{Pa}$               & \multirow{4}{*}{Prognostic, Instantaneous}\\
& 2-Meter Temperature                        & $\mathrm{K}$                & \\
& 10-Meter Zonal Wind                        & $\mathrm{m \cdot s^{-1}}$   &\\
& 10-Meter Meridional Wind                   & $\mathrm{m \cdot s^{-1}}$   &\\
\midrule
\multirow{12}{*}{Flux form\textsuperscript{b}} & Total Precipitation          & $\mathrm{m}$                & \multirow{10}{*}{Diagnostic, Cumulative} \\
& Evaporation                                & $\mathrm{m}$                & \\
& Top-of-atmosphere Net Solar Radiation      & $\mathrm{J \cdot m^{-2}}$   &\\
& Outgoing Longwave Radiation                & $\mathrm{J \cdot m^{-2}}$   &\\
& Surface Net Solar Radiation                & $\mathrm{J \cdot m^{-2}}$   &\\
& Surface Net Longwave Radiation             & $\mathrm{J \cdot m^{-2}}$   &\\
& Surface Net Sensible Heat Flux           & $\mathrm{J \cdot m^{-2}}$   &\\
& Surface Net Latent Heat Flux             & $\mathrm{J \cdot m^{-2}}$   &\\
\cmidrule(lr){2-4}
& Top-of-atmosphere Incident Solar Radiation & $\mathrm{J \cdot m^{-2}}$   & Input-only, Cumulative \\
\midrule
\multirow{4}{*}{Others}                  & Sea-ice Cover                              & n/a                         & Input-only, Instantaneous \\
& Geopotential at the Surface                & $\mathrm{m^2 \cdot s^{-2}}$ & Input-only, Static \\
& Land-sea Mask                              & n/a                         & Input-only, Static \\
& Soil Type                                  & n/a                         & Input-only, Static \\
\specialrule{1.5pt}{3pt}{0pt}
\end{tabularx}
\end{center}
\textsuperscript{a} Specific total water is the combination of specific humidity, cloud liquid water content, and rainwater content.\\
\textsuperscript{b} Flux form variables are accumulated every 6 hours. Downward flux is positive.
\end{table}

\begin{figure}
    \centering
    \includegraphics[width=\columnwidth]{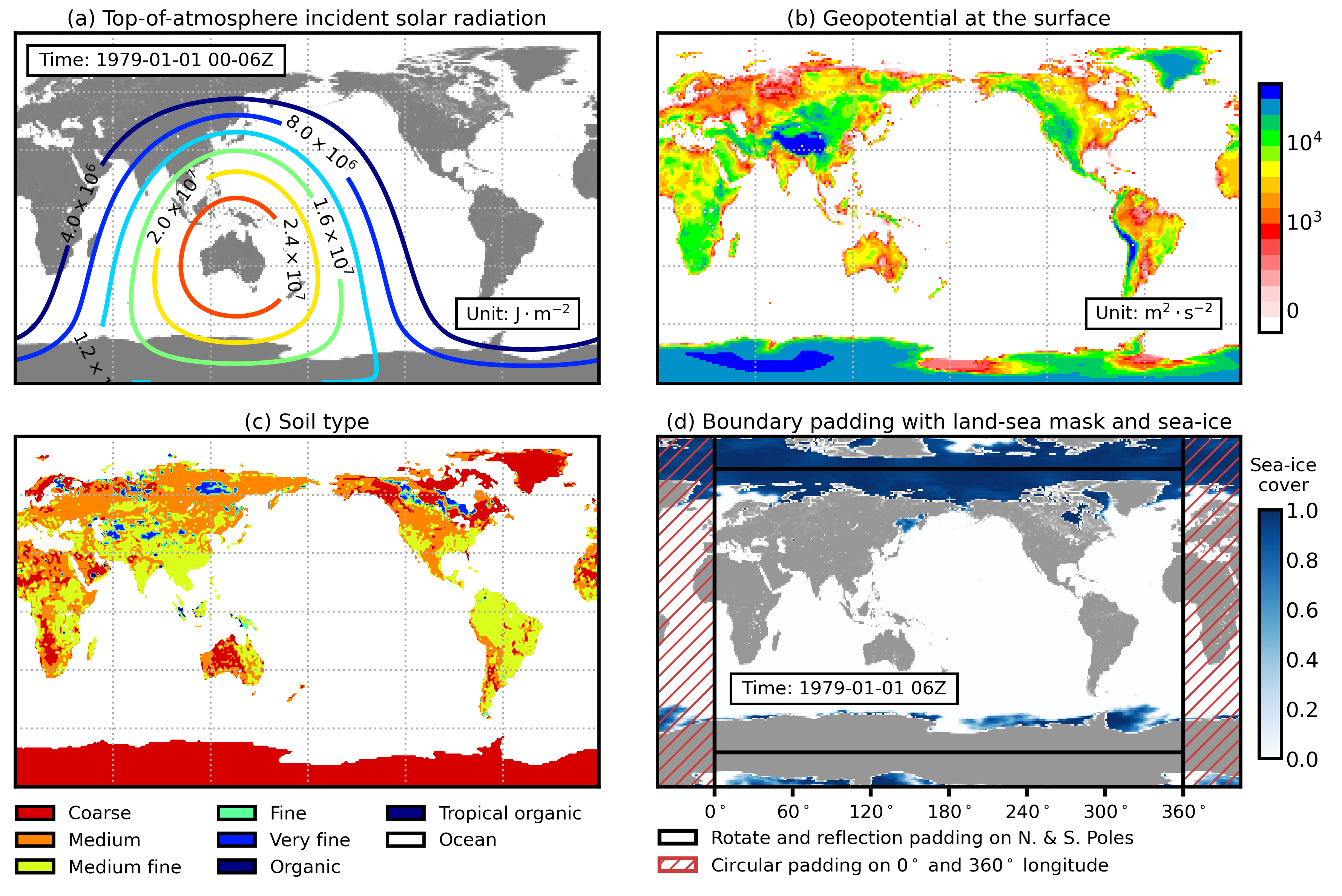}
    \caption{Example input-only variables of (a) Top-of-atmosphere incident solar radiation, accumulated in 0000-0600 UTC, 1 January 1979. (b) Geopotential at the surface. (c) Soil type, and (d) Land-sea mask with sea-ice cover on 0600 UTC, 1 January 1979. Hatched areas in (d) are values created from the boundary padding.}
    \label{fig1}
\end{figure}

The primary dataset used in this study is ERA5, a global reanalysis that has been extensively utilized in the development of AIWP models. It is produced using the Integrated Forecasting System (IFS) version Cy41r2 at ECMWF, incorporating state-of-the-art model physics, dynamics, and data assimilation \cite{hersbach2020era5}.

A wide range of variables are selected to represent the state of the atmosphere and its interface with other earth system components (Table \ref{tab1}). Many flux form variables are applied---this is the first time that a complete set of moisture, energy, and radiation fluxes at the top of the atmosphere and the surface are integrated into AIWP models. They  play critical roles in estimating the sources and sinks of the atmospheric moisture and energy budgets. Geopotential at the surface is an indicator of the terrain elevation (Figure~\ref{fig1}b). For specific total water, we consider water vapor and liquid-phase moisture only. We found that solid-phase moisture can introduce many outliers in the upper atmosphere. 

The ERA5 dataset used in this study is obtained from the NSF NCAR, Research Data Archive \cite{ecmwf2019era5}, and the Google Research, Analysis-Ready, Cloud Optimized (ARCO) ERA5 \cite{carver2023arcoera5}. All variables in Table \ref{tab1} are re-gridded to 1.0$^\circ$ grid spacing, 181-by-360 horizontal grid cells, using conservative interpolation. For vertical dimensions, 1, 50, 150, 200, 250, 300, 400, 500, 600, 700, 850, 925, 1000 hPa levels are selected. The pre-processed ERA5 variables, as described above, are prepared as 6-hourly datasets spanning 1 January 1979 to 31 December 2021 following the validity time convention of ECMWF. Instantaneous variables are selected based on the ending time of every 6 hours, whereas flux form variables are accumulated within each 6-hour time window (e.g., Figure~\ref{fig1}a). The 6-hour dataset is then divided into three parts, with 1979-2018 used for model training, 2019 used for validation, and 2020-2021 used for verification.

The ERA5 climatology is involved in the verification of total precipitation. It is also used as a climatology reference for some of the flux form variables. The ERA5 climatology used in this study is consistent with Weatherbench2, which is calculated from the 30-year period of 1990-2019 using sliding windows of 61 days with Gaussian weighting \cite{rasp2024weatherbench}.

\subsection{Data pre-processing}\label{sec22}

All ERA5 variables in Table \ref{tab1}, except sea-ice cover, land-sea mask, and soil type, are normalized using z-score. The required mean and standard deviation are computed within the training set of 1979-2018. For the remaining variables, soil type is rescaled from categorical integer values to 0.0-1.0 float numbers. The land-sea mask and the sea-ice cover are combined, with pure land, ocean, and sea-ice having float numbers of 1.0, 0.0, and -1.0, respectively (e.g., Figure~\ref{fig2}c and d).

\section{Methods}\label{sec3}

\subsection{FuXi with physical constraints}\label{sec31}

\begin{figure}
    \centering
    \includegraphics[width=\columnwidth]{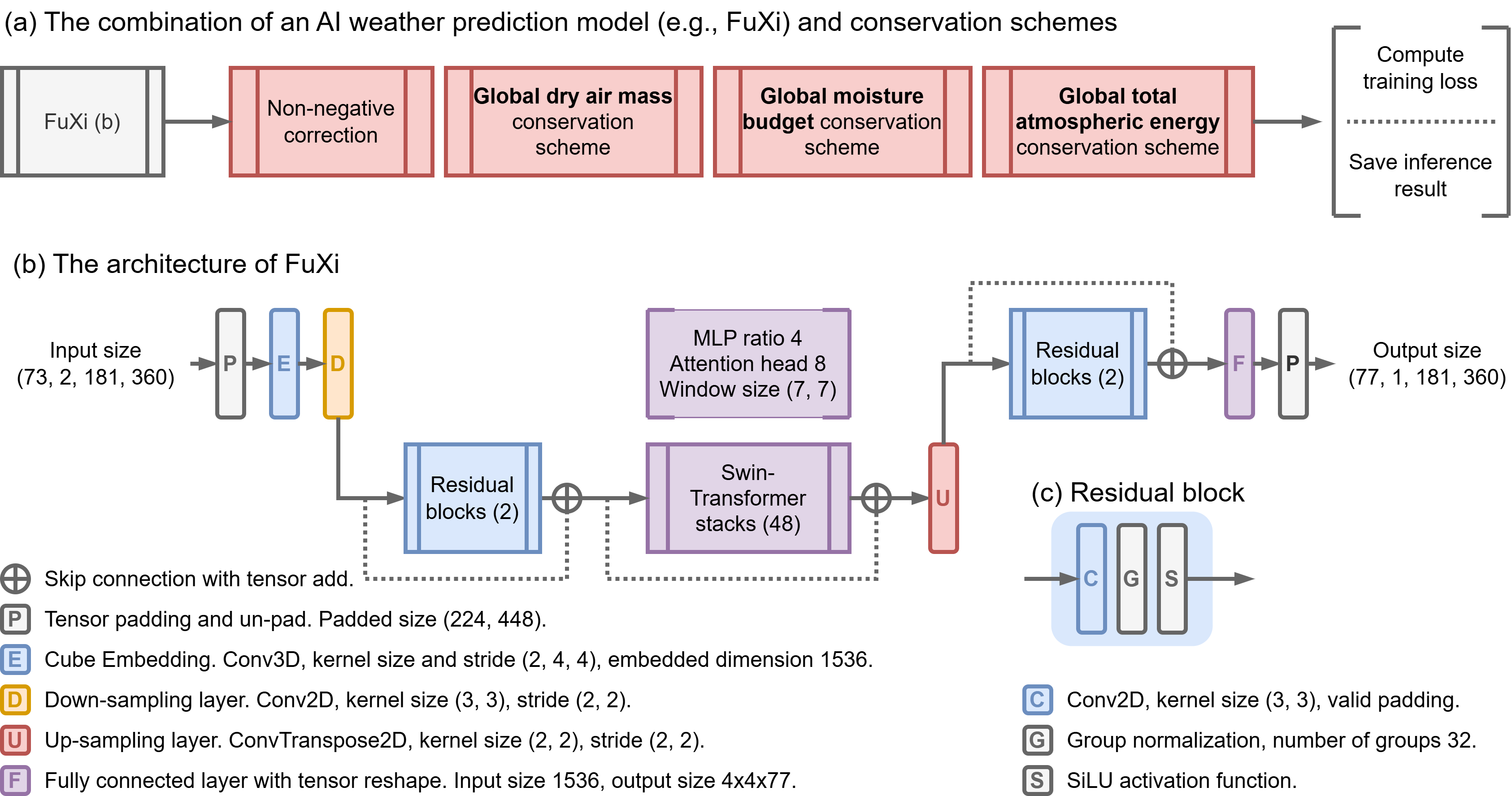}
    \caption{(a) The arrangement of conservation schemes during the training and inference of FuXi. (b) The architecture of FuXi with the I/O sizes of this study. ``(48)'' means repeat 48 times. ``MLP ratio 4'' means the number of Multi-Layer Perceptron (MLP) neurons is 4$\times$ the number of embedded dimensions. (c) The architecture of the residual block.}
    \label{fig2}
\end{figure}

FuXi \cite{chen2023fuxi}, a state-of-the-art AIWP model, is selected as an example to implement conservation schemes. The base architecture of FuXi is Swin-Transformer, which has been adopted by a wide range of AIWP models \cite<e.g.>{willard2024analyzing,nguyen2023scaling}, making FuXi an ideal candidate for showcasing the integration of physical constraints. FuXi has also achieved top-tier scores in Weatherbench2 \cite{rasp2024weatherbench}, which means that its pure AIWP model version can be used as a strong baseline for evaluating the improvements introduced by the conservation schemes.

The architecture and hyperparameters of FuXi follow its original design [Figure~\ref{fig2}b; \citeA{chen2023fuxi}]. The cube embedding layer employs patch sizes of (2, 4, 4) for the time, latitude, and longitude dimensions, respectively, with an embedded dimension of 1536. The U-Transformer consists of convolutional-layer-based tensor resampling, residual blocks, and 48 Swin-Transformer V2 blocks \cite{liu2022swin}. The Swin-Transformer blocks are configured with eight attention heads, window sizes of (7, 7), and MLP ratios of 4.

A few changes are made based on the original FuXi design. (1) Model cascading is not applied. It is determined that training multiple FuXi for different forecast lead time groups is unnecessary for examining the behavior of the conservation schemes. (2) Boundary padding is applied; it performs circular padding on 0$\mathrm{^\circ}$ and 360$\mathrm{^\circ}$ longitude and reflection padding with 180$\mathrm{^\circ}$ rotation for the North and South Poles (see hatched area in Figure~\ref{fig1}d). Here, boundary padding adjusts the input sizes from (181, 360) to (224, 448), ensuring that the input tensor fits with the (7, 7) attention windows of the U-Transformer without additional interpolations. (3) Spectral normalization is applied to all trainable layers of the model; it stabilizes the model during training \cite<e.g.>{miyato2018spectral,lin2021spectral}. Changes above were implemented similarly in \citeA{schreck2024community} (i.e., referred to as ``CREDIT-FuXi'' in theirs) and reported competitive performance.

Conservation schemes are applied after the FuXi output layer (Figure~\ref{fig1}a). The purposes and methods of the conservation schemes are described below, with the adjusted variables emphasized.

\begin{enumerate}
    \item Nonnegative correction: AIWP models can produce negative values. For all nonnegative variables (e.g., specific total water, total precipitation), their negative raw outputs will be corrected to zero.
    
    \item Global dry air mass conservation scheme. At each forecast step, \textit{specific total water} is corrected to ensure that the global dry air mass content stays the same as in the initial condition.
    
    \item Global moisture budget conservation scheme. At each forecast step, \textit{total precipitation} is corrected to balance the global sum of the total precipitable water tendency, derived from specific total water, and the accumulated net flux of total precipitation and evaporation over the previous 6-hour period. 
    
    \item Global total atmospheric energy conservation scheme. Global energy conservation is defined as the balance between the tendency of total atmospheric energy and net energy fluxes on the top of the atmosphere and the surface. At each forecast step, \textit{air temperature} is corrected to balance the global sum of total atmospheric energy tendencies and the energy sources and sinks over the past 6-hour period.
\end{enumerate}

The variable names used above align with those listed in Table \ref{tab1}. Corrections to specific total water, total precipitation, and air temperature are applied using multiplicative ratios across all grid cells. For specific total water, corrections are limited to pressure levels below 600 hPa, as the moisture content in the upper atmosphere is significantly lower, and enforcing corrections at higher levels would lead to strong outlier effects within the z-score space and impacts training stability.

Note that the conservation schemes above must be applied in the specified order to achieve the desired performance. Further technical details are available in Appendix \ref{A1}.

\subsection{Experiment design and model training}\label{sec32}

Two FuXi configurations are prepared for the experiment: ``FuXi-base'', the baseline configuration without using conservation schemes, and ``FuXi-physics'', the main configuration with conservation schemes. By comparing the performance of the two, the benefits of using conservation schemes can be identified. 

To provide a reference for performance, the ECMWF IFS high-resolution forecast (IFS-HRES) is included as an NWP baseline. IFS-HRES is widely recognized as the best operational medium-range weather forecasting system and is commonly used as a benchmark in AIWP studies. IFS-HRES forecasts initialized on 0000 and 0012 UTC (hereafter, ``00Z'' and ``12Z'', respectively) are collected from the Weatherbench 2 data archive \cite{rasp2024weatherbench} and interpolated to the 1.0$^\circ$ grid spacing. Hereafter, this 1.0$^\circ$ interpolated version of IFS-HRES is referred to as ``IFS-HRES'' directly. 

It is important to note, however, that because of the ongoing improvements of IFS, the operational model is expected to be more skillful. As such, the main focus of the experiment is the comparison between FuXi-base and FuXi-physics, with IFS-HRES serving as a sanity check. Its participation ensures that the implementation of FuXi in this study is valid. We have no intention of downplaying the importance and forecasting skills of IFS-HRES.

Training procedures for the two FuXi configurations largely follow \citeA{lam2023learning} and \citeA{chen2023fuxi}. For single-step training, a warm-up stage is prepared first; it has 10 epochs, with 100 batches per epoch and 32 samples per batch. After, 170 full epochs are trained, with roughly 1800 batches per epoch. The AdamW optimizer is used with weight decay of $3\times 10^{-6}$ \cite<c.f.>{bi2023accurate}, an initial learning rate of $10^{-3}$, and the (half) cosine-annealing learning rate schedule. For multi-step training, a fixed learning rate of $3\times 10^{-7}$ is applied over 12 full epochs. The number of iterative steps increases per epoch from 2 to 12, corresponding to forecast lead times of 12 to 72 hours. Training loss is calculated using mean squared error with the inverse variance weights and per-variable-level weights. Technical details of the training objectives are summarized in Appendix \ref{A2}.

The two FuXi configurations are trained on 32 NVIDIA A100 GPUs using Pytorch \cite{paszke2019pytorch}. Under fully sharded data parallel \cite{zhao2023pytorch} with activation checkpointing, the speed of single-step training is roughly 50 minutes per epoch for FuXi-base and 60 minutes per epoch for FuXi-physics. The speed of multi-step training varies by the number of steps. For 11-step (72-hour) training, it is roughly 14-16 hours per epoch for the two models, with FuXi-physics being slightly slower.  

\subsection{Verification methods}\label{sec33}

The verification of this study covers two parts: (1) The conservation of global dry air mass, moisture budget, and total atmospheric energy is verified. The goal of this verification is to confirm that the conservation schemes work as expected. (2) Deterministic verifications are conducted to examine the forecast skill. This part helps identify the performance difference of incorporating conservation schemes in AIWP models.

For the verification of conservation properties, key components of the conservation relationships, as well as residual terms that violate the conservation, will be computed from the two FuXi runs. While forecasts that violate conservation laws are not necessarily inaccurate, consistent violation of global mass and energy conservation can cause the model climate to drift and potentially become unstable. FuXi-physics will be considered successful if it demonstrates the ability to conserve global mass and energy in its iterative forecasts.

Two ERA5 datasets are involved in this verification: the pre-processed 1.0$^\circ$ ERA5, as described in Section \ref{sec21}, and the original 0.25$^\circ$ ERA5 with 37 pressure levels and all moisture components. Aside from the verifications of the two FuXi runs, comparisons of the two ERA5 datasets would serve as a proof, showing that the pre-processed 1.0$^\circ$ ERA5 dataset of this study generally preserved the global mass and energy conservation properties as presented in the original ERA5. Note that ERA5 is a reanalysis dataset, and its most commonly used pressure-level version does not conserve mass and energy strictly \cite<e.g.>{trenberth2018applications,mayer2021consistency}. Therefore, the conservation residuals of ERA5 are not zero. This issue will be discussed later in Section \ref{sec53}.

The deterministic verification of this study primarily follows that of Weatherbench2 \cite{rasp2024weatherbench}. The performance of the two FuXi runs is examined using the pre-processed 1.0$^\circ$ ERA5 as the verification target. The performance of IFS-HRES is also verified and serves as a reference.

Most variables are verified using Root Mean Squared Error (RMSE) and its increments relative to the IFS-HRES. Global averaged RMSE values are calculated using latitude-based weighting \cite<e.g.>{schreck2024community,rasp2024weatherbench}. For instantaneous variables, verification is conducted every 6 hours; for cumulative variables, they are aggregated from 6 hours to daily totals first. The RMSE maps on day 10 are also analyzed to identify the spatial signals of performance differences.

Many flux form variables (see Table \ref{tab1}) are not part of the Weatherbench2 \cite{rasp2024weatherbench}, and they are not covered fully by any other AIWP models. To be consistent with \citeA{rasp2024weatherbench}, the RMSE verification of flux form variables will not have IFS-HRES involved. The purpose of this part is to ensure that the two FuXi runs can produce a reasonable amount of mass and energy fluxes. That said, it is not about benchmarking flux form variables against IFS-HRES. The RMSE computed from the ERA5 climatology (see Section \ref{sec21}) is provided as a reference.

Total precipitation is verified using Stable Equitable Error in Probability Space (SEEPS). SEEPS is a verification metric designed to quantify the accuracy of precipitation forecasts. It converts the raw precipitation amounts to dry, light, and heavy precipitation categories using climatology-based thresholds \cite{rodwell2010new}. SEEPS is better than RMSE for verifying precipitation because it penalizes spatially smoothed forecasts correctly. Note that the verification target of total precipitation is ERA5. Here, the ERA5-based verification of total precipitation is sufficient to distinguish the two FuXi runs. As it will be discussed later, the caveats of ERA5 do not deny the demonstrated benefit of conservation schemes, which is the reduction of drizzle bias over the entire domain. That said, the ERA5 is not used to show the RMSE difference in a specific area. This issue will also be discussed further in Section \ref{sec512}.
 
The global kinetic energy and potential temperature energy spectrum of forecasts are verified on zonal wavenumbers using spherical harmonic transforms. The zonal energy spectrum measures the energy transfer relationships across scales, which quantifies the effective resolution of a given forecast. As pointed out in \citeA{rasp2024weatherbench}, this verification exposes the extent of over smoothing in AIWP models.

For deterministic verifications, the two FuXi runs will produce forecasts initialized on 00Z and 12Z in 2020-2021 for up to 10 days. For the verification of conservation properties, the two FuXi runs will produce 00Z forecasts for up to 15 days. Verifications from all initializations in 2020-2021 are averaged either spatially with latitude weighting or temporally to produce results. Technical details are summarized in Appendix \ref{A3}.
\section{Results}\label{sec4}

\subsection{The conservation of global mass and energy}\label{sec41}

\begin{figure}
    \centering
    \includegraphics[width=\columnwidth]{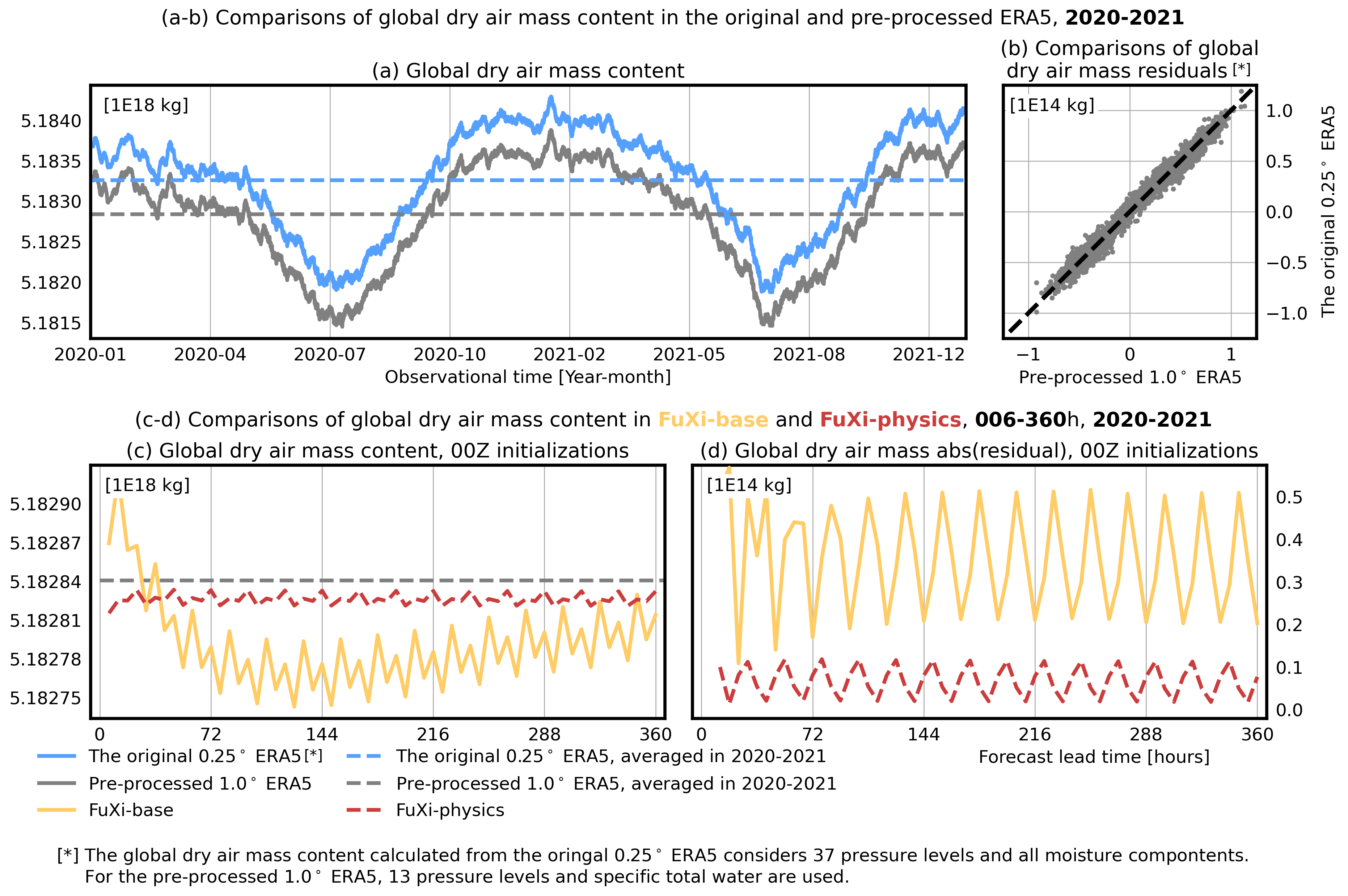}
    \caption{Verification of global dry air mass conservation in 2020-2021. (a) Global dry air mass content computed from the original 0.25$^\circ$ ERA5 (blue) and the pre-processed 1.0 $^\circ$ ERA5 (gray). (b) Quantitative comparisons of global dry air mass conservation residuals. (c) Global dry air mass content of FuXi-base (orange solid line) and FuXi-physics (red dash line) averaged from all 00Z initializations and up to day-15 forecasts. (d) As in (c), but for the absolute values of global dry air mass conservation residuals. Panels (a)-(c) have scales of $10^{18}$ kg; panel (d) has the scale of $10^{14}$ kg.}
    \label{fig3}
\end{figure}

Figure~\ref{fig3} provides an assessment of the global dry air mass conservation in the ERA5 datasets and the two FuXi runs. The total amount of global dry air mass is expected to be constant regardless of time, so any remaining tendency is a conservation residual. ERA5 conserved the global dry air mass well, with the analyzed amount staying around $\mathrm{5.18\times 10^{18}}$ kg (Figure~\ref{fig3}a). This value does not agree with other more accurate analyses [e.g., $\mathrm{5.13\times 10^{18}}$ kg in \citeA{trenberth2005mass}] that considers surface pressure and orography (here we consider constant pressure levels from 1000 to 1 hPa), but it is generally stable over time with residuals on the scale of $\mathrm{10^{14}}$ kg (Figure~\ref{fig3}b). 

A small dry air mass content difference is found between the 1.0$^\circ$ pre-processed ERA5 and the 0.25$^\circ$ original ERA5 (c.f. blue and gray lines in Figure~\ref{fig3}a). However, their conservation residuals are almost identical (Figure~\ref{fig3}b). This shows that the impact of ERA5 pre-processing steps is generally minor.

Taking the pre-processed 1.0$^\circ$ ERA5 as training targets, FuXi-base inherited a similar amount of dry air mass conservation residuals on the scale of $\mathrm{10^{14}}$ kg. Notably, its total amount of dry air mass varies with forecast lead time, with a decrease from 6 to 72 hours and then an increase from 72 hours to 360 hours. The variation is small, but since it is a forecast and the variation has a specific pattern, it still raises concerns that pure AIWP models, such as FuXi-base, may lose the ability to maintain a stable amount of global dry air mass after certain numbers of iterative steps. On the contrary, FuXi-physics, equipped with the conservation schemes, produced a stable amount of global dry air mass, and its conservation residual is 80-95\% lower than that of the FuXi-base (Figure~\ref{fig3}d). Note that the conservation schemes cannot suppress residual terms to strictly zero. This is due to the limitation of numerical precisions. For more details, see Appendix \ref{A15}.

\begin{figure}
    \centering
    \includegraphics[width=\columnwidth]{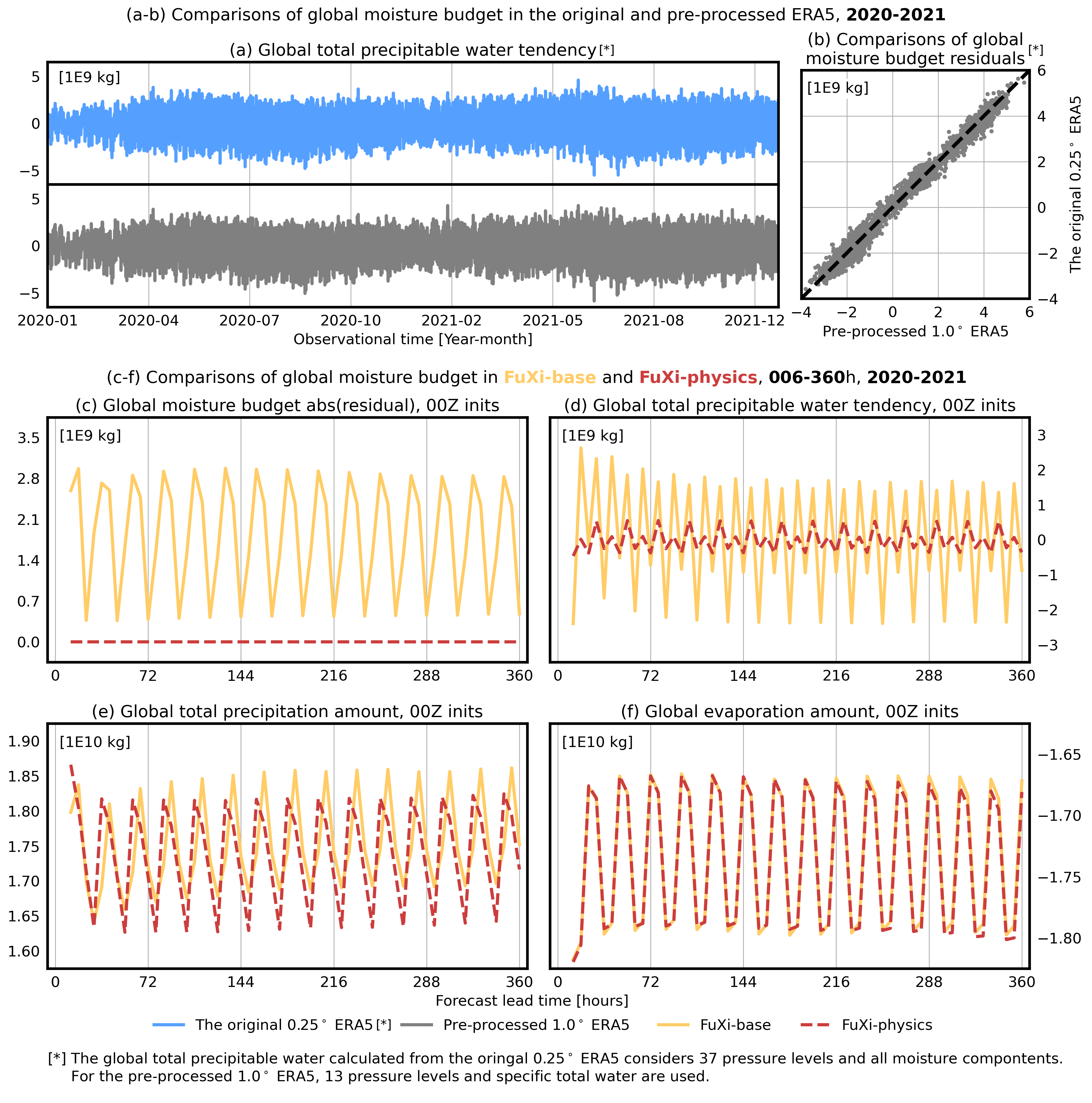}
    \caption{Verification of global moisture budget conservation in 2020-2021. (a) Global total precipitable water tendencies computed from the original 0.25$^\circ$ ERA5 (blue) and the pre-processed 1.0 $^\circ$ ERA5 (gray). (b) Quantitative comparisons of global moisture budget conservation residuals. (c) The absolute values of global moisture budget conservation residuals of FuXi-base (orange solid line) and FuXi-physics (red dash line) averaged from all 00Z initializations and up to day-15 forecasts. (d) As in (c), but for global total precipitable water tendencies. (e) As in (c), but for global total precipitation amount. (f) As in (c), but for global total evaporation amount. Panels (a)-(d) have scales of $10^{9}$ kg; panels (e) and (f) have scales of $10^{10}$ kg.}
    \label{fig4}
\end{figure}

Figure~\ref{fig4} examines the conservation of global moisture budget, i.e., the balance between the global precipitable water tendencies and the net evaporation/total precipitation fluxes. The original 0.25$^\circ$ ERA5 and the pre-processed 1.0$^\circ$ ERA5 are verified to have highly similar tendency terms and moisture budget residuals. This further confirms the effectiveness of the pre-processed 1.0$^\circ$ ERA5 on preserving the mass conservation signals of the original ERA5 (Figure~\ref{fig4}a and b).

The two FuXi runs showed notable differences in the conservation of moisture budget. The residual of FuXi-base is on the scale of $\mathrm{10^{9}}$ kg (Figure~\ref{fig4}c), comparable to that of ERA5, but clearly larger than FuXi-physics. Two factors are responsible for explaining such differences: 

\begin{enumerate}
    \item The oscillation of global total precipitable water tendency in the FuXi-base is too high, and the magnitude ranges from -2.5 to 1.5$\times 10^9$ kg. That said, the tendency term is not stable over time, and it is not zero-centered (orange lines in Figure~\ref{fig3}c and Figure~\ref{fig4}d). 
    \item The two FuXi runs produced highly similar global evaporation amounts, but the global total precipitation amount from FuXi-base is roughly $\mathrm{0.5\times 10^{9}}$ kg higher than that of the FuXi-physics (Figure~\ref{fig4}e). 
\end{enumerate}

The first reason shows the benefit of dry air mass conservation scheme---it not only maintains a stable global dry air mass amount but also suppresses the oscillation of total precipitable water tendency, thus providing additional support to the simulation of global moisture budget. It is also one of the reasons why conservation schemes must run in a specific order. For the second reason, the FuXi-base likely overestimated the global total precipitation amount due to its drizzle bias. This part will be elaborated later in the SEEPS verification section. 

FuXi-physics is verified to conserve its moisture budget. Its conservation residual is nearly zero, verified to be on the scale of $\mathrm{10^{1}}$-$\mathrm{10^{3}}$ kg. Its global total precipitable water tendency and total precipitation amount are also more stable compared to that of FuXi-base. All evidence points to successful implementation of the conservation schemes.

\begin{figure}
    \centering
    \includegraphics[width=\columnwidth]{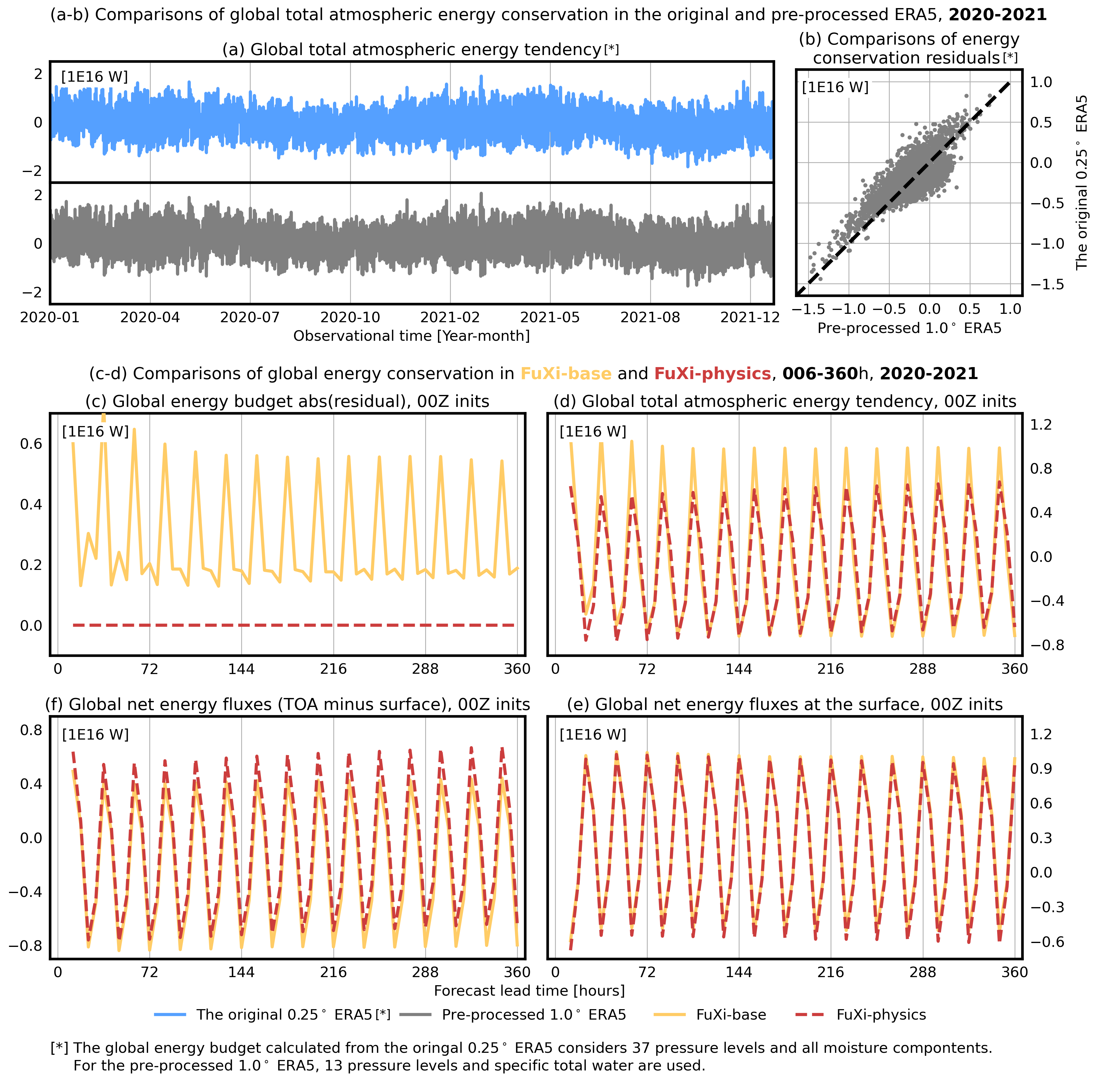}
    \caption{Verification of global total atmospheric energy conservation in 2020-2021. (a) Global total atmospheric energy tendencies are computed from the original 0.25$^\circ$ ERA5 (blue) and the pre-processed 1.0 $^\circ$ ERA5 (gray). (b) Quantitative comparisons of global energy budget conservation residuals. (c) The absolute values of global energy budget conservation residuals of FuXi-base (orange solid line) and FuXi-physics (red dash line) averaged from all 00Z initializations and up to day-15 forecasts. (d) As in (c), but for global total atmospheric energy tendencies. (e) As in (c), but for global net energy fluxes. (f) As in (c), but for global net energy fluxes at the surface. All panels have scales of $10^{16}$ W.}
    \label{fig5}
\end{figure}

Global total atmospheric energy conservation is verified in Figure~\ref{fig5}. This verification is based on the balance between the global total atmospheric energy tendency and energy fluxes on the top of the atmosphere and the surface. The original 0.25$^\circ$ ERA5 and the pre-processed 1.0$^\circ$ ERA5 have similar total atmospheric energy tendencies, with a few outlier points found from the energy conservation residuals (Figure~\ref{fig5}a and b). These outliers are mostly related to the vertical level coarsening of the pre-processed 1.0$^\circ$ ERA5. By reducing its vertical levels from 37 to 13, the pre-processed ERA5 may lose some ability to capture the variation of air temperature in the upper atmosphere. This can be translated to a small difference in the estimation of thermal energy. Despite this issue, the energy conservation residuals of the two ERA5 datasets stayed mostly along the diagonal reference line, showing that the pre-processed 1.0$^\circ$ ERA5 captured most of the energy conservation patterns from its original 0.25$^\circ$ version.

The two FuXi runs exhibited different energy conservation behaviors. FuXi-base produced energy conservation residuals on the scale of $\mathrm{10^{16}}$ W. For FuXi-physics, the residual is way smaller, verified to be on the scale of $\mathrm{10^{9}}$ W (Figure~\ref{fig5}c). The two FuXi runs agree well on their energy sources and sinks (Figure~\ref{fig5}e and f), and their energy conservation difference is mainly explained by the total atmospheric energy tendency. For FuXi-physics, the tendency term ranges from -0.5 to 0.5 $\mathrm{10^{16} W}$ and is centered around zero. For FuXi-base, although its tendency term shares the same diurnal cycle with that of the FuXi-physics, its value range is [-0.5, 1.0] $\mathrm{10^{16} W}$, drifted away from the zero center. This can lead to an overestimation of energy stored in the atmosphere and cause an imbalance. Comparing the FuXi-base verification in Figure~\ref{fig5}c and d, clear connections can be found between the diurnal peaks of conservation residuals and the peaks of energy tendency.

In summary, the conservation schemes are verified to work correctly. FuXi-physics, equipped with the conservation schemes, performed clearly better than FuXi-base with lower mass and energy conservation residuals and more stable conservation properties (e.g., global dry air mass content, total precipitation amount, total atmospheric energy tendency) during its iterative forecasts. For the FuXi-base, although not supported by the conservation schemes, its residual terms stay at roughly the same level as in the ERA5 up to day 15. 

The conservation behaviors of the original 0.25$^\circ$ ERA5 and the pre-processed 1.0$^\circ$ ERA5 are also compared. The verification results point to a good agreement, indicating that the ERA5 pre-processing steps (i.e., vertical level coarsening, interpolation, and the simplification of moisture representation) generally preserved its conservation properties. Thus, the pre-processed 1.0$^\circ$ ERA5 is qualified as training and verification targets.

\subsection{Deterministic verification}\label{sec42}

\subsubsection{Total precipitation}\label{sec424}

\begin{figure}
    \centering
    \includegraphics[width=\columnwidth]{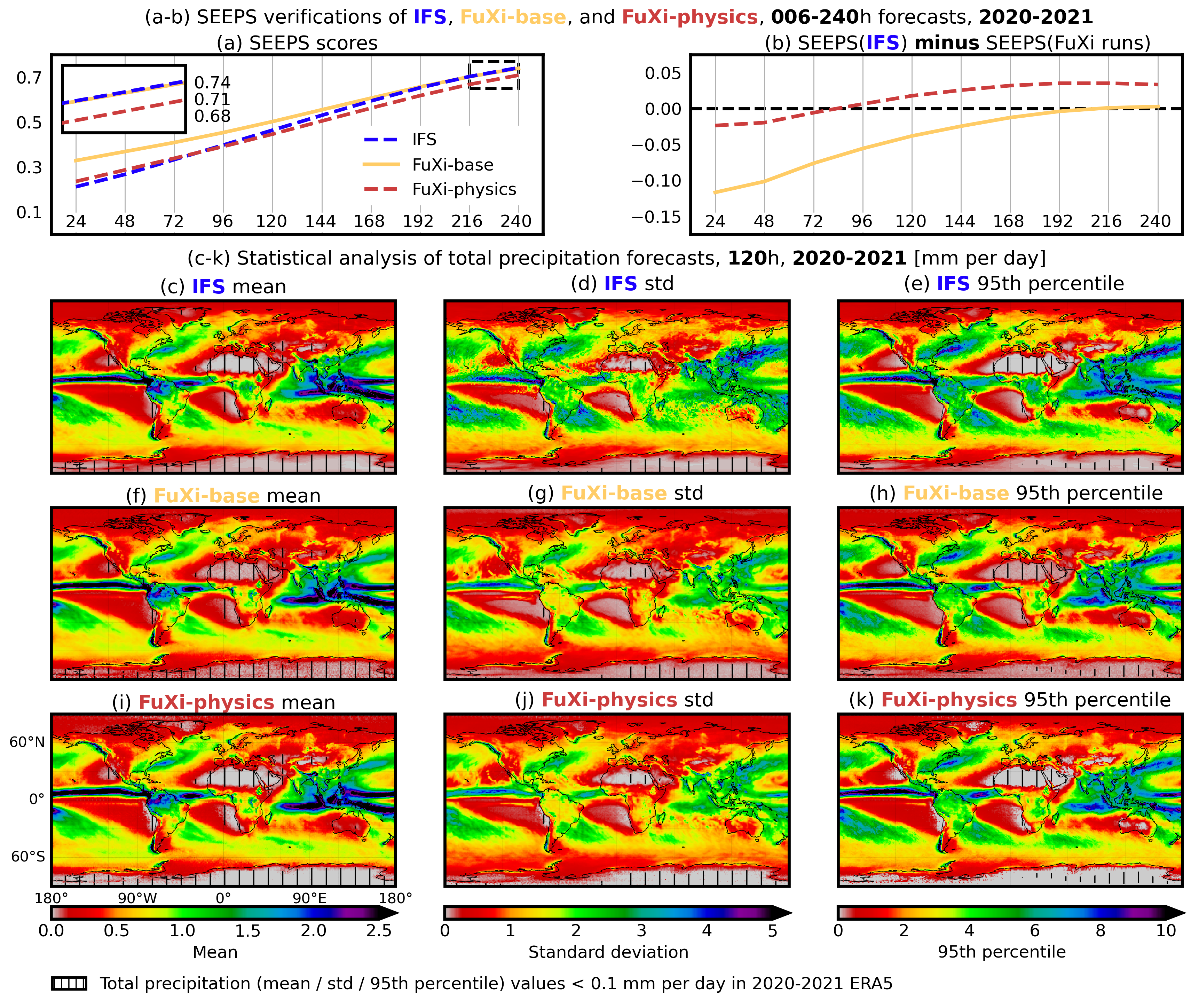}
    \caption{SEEPS verification of accumulated daily total precipitation in IFS-HRES (blue dashed line), FuXi-base (orange solid line), and FuXi-physics (red dashed line) for all 00 and 12Z initializations in 2020-2021. (a) SEEPS by forecast lead times from 06 to 240 hours. Lower is better. The smaller box indicates a zoom-in. (b) The SEEPS of IFS-HRES minus the two FuXi runs. Positive and higher means better than IFS-HRES. (c-e) the gridpoint-wise mean, standard deviation (``std''), and 95th percentile of IFS-HRES day 5 forecasts. Hatch highlights dry regions where the value of each statistic is below 0.1 mm per day in the ERA5. (f-h) As in (c-e), but for FuXi-base. (j-l) As in (c-e), but for FuXi-physics.}
    \label{fig11}
\end{figure}

Figure~\ref{fig11} shows the daily total precipitation verification results using SEEPS. FuXi-base is verified to be worse than the IFS-HRES. For 24 and 48-hour forecasts, its performance downgrade is roughly 20\%. This result is expected and aligns with other AIWP models verified in \citeA{rasp2024weatherbench}. AIWP models are typically less competitive in SEEPS verifications than IFS-HRES due to their blurriness, which translated to the overestimation of light precipitation in categorical scores. FuXi-physics is verified to be clearly better than FuXi-base. Its SEEPS scores are slightly worse than IFS-HRES for up to day 3, and then outperform IFS-HRES at longer forecast lead times with roughly 5-7\% improvements (Figure~\ref{fig11}a, b).

Statistical analysis of total precipitation forecasts is conducted to investigate the difference between FuXi-base and FuXi-physics. FuXi-base is found to produce large amounts of light precipitation incorrectly in climatologically dry areas, such as the South Pole, Sahal Desert, and West African Coast, defined using the 0.1 mm per day threshold in the ERA5 (i.e., hatched areas in Figure~\ref{fig11}f-h). Such overestimation of drizzles is persistent and systematic because they can be found in mean, standard deviation, and 95th percentiles, all in the same locations (i.e., areas with red color shades in Figure~\ref{fig11}f-h).  

FuXi-physics, equipped with the conservation schemes, successfully avoided such drizzle bias. It shows a clearly better separation between the dry and light precipitation areas. This can be verified by comparing (1) areas with gray colored shade and hatches in Figure~\ref{fig11}i-k for FuXi-physics, and (2) areas with red color shades and hatches in Figure~\ref{fig11}f-h for FuXi-base. A good dry-to-light precipitation separation is a large improvement against the blurriness problems of pure AIWP models, and it is encouraged in SEEPS verification. In addition, the performance of FuXi-physics is on the same level as IFS-HRES and FuXi-base for the spatial distribution and variability of extreme precipitation events (c.f. Figure~\ref{fig11}d, g, j and e, h, k), which ultimately led to better total precipitation verification results.

\subsubsection{Upper-air and near-surface prognostic variables}\label{sec421}

\begin{figure}
    \centering
    \includegraphics[width=\columnwidth]{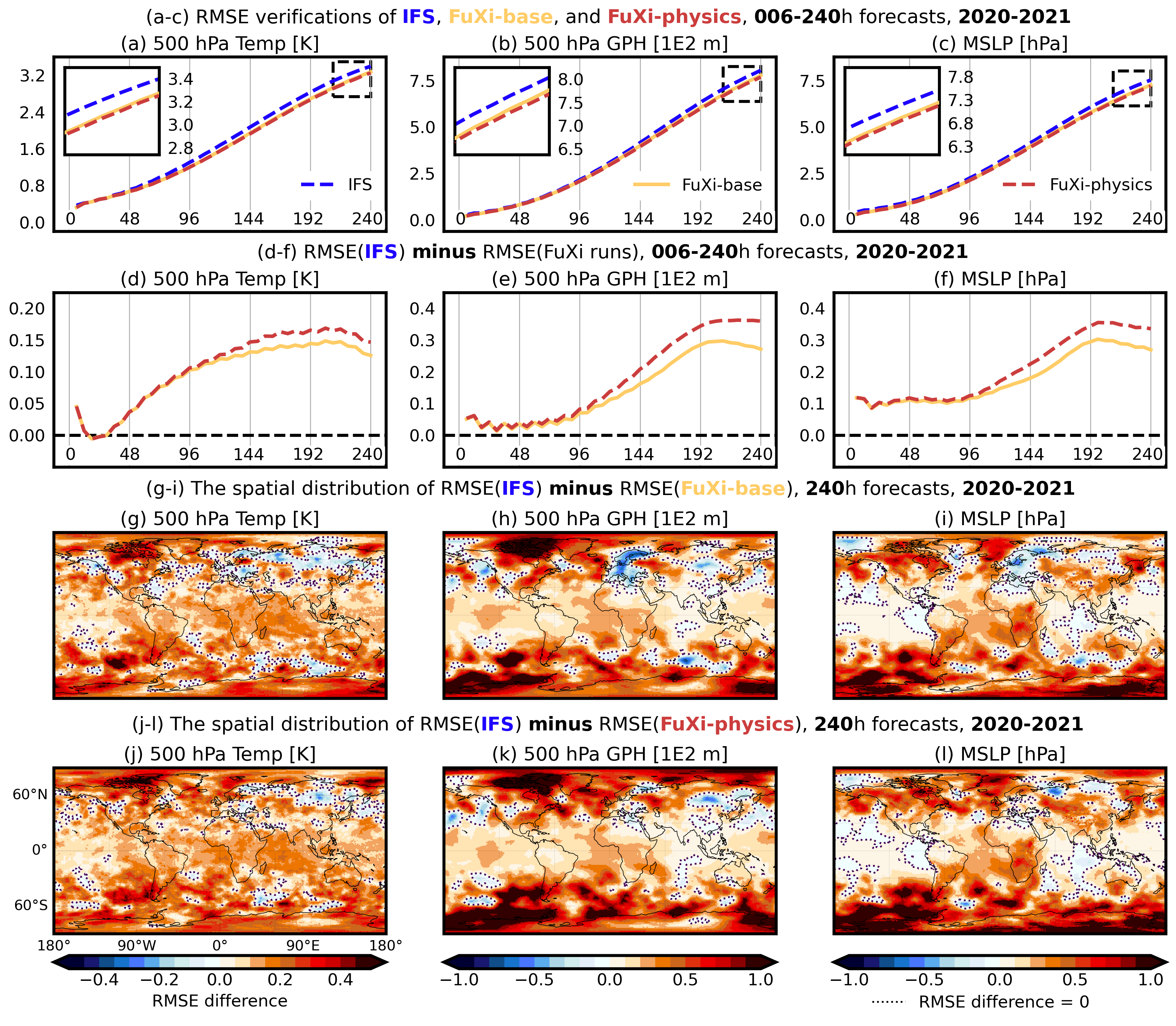}
    \caption{RMSE verifications of IFS-HRES (blue dashed line), FuXi-base (orange solid line), and FuXi-physics (red dashed line) for all 00 and 12Z initializations in 2020-2021. (a-c) Domain-weighted averaged RMSE by 6-hourly forecast lead times for 500 hPa air temperature (``Temp''), geopotential height (``GPH''), and mean sea level pressure (MSLP), respectively. Lower is better. The smaller box indicates a zoom-in. (d-f) The RMSE of IFS-HRES minus the two FuXi runs. Positive and higher means better than IFS-HRES, vice versa. (g-i) The spatial distribution of the RMSE of IFS-HRES minus FuXi-base on day 10 forecasts. Higher/red colored shades are better. Dashed lines indicate that the RMSE difference is zero. (j-l) as in (g-i), but for FuXi-physics. Note that panels (b), (e), (h), and (k) have scales of $10^{2}$ m.}
    \label{fig6}
\end{figure}

\begin{figure}
    \centering
    \includegraphics[width=\columnwidth]{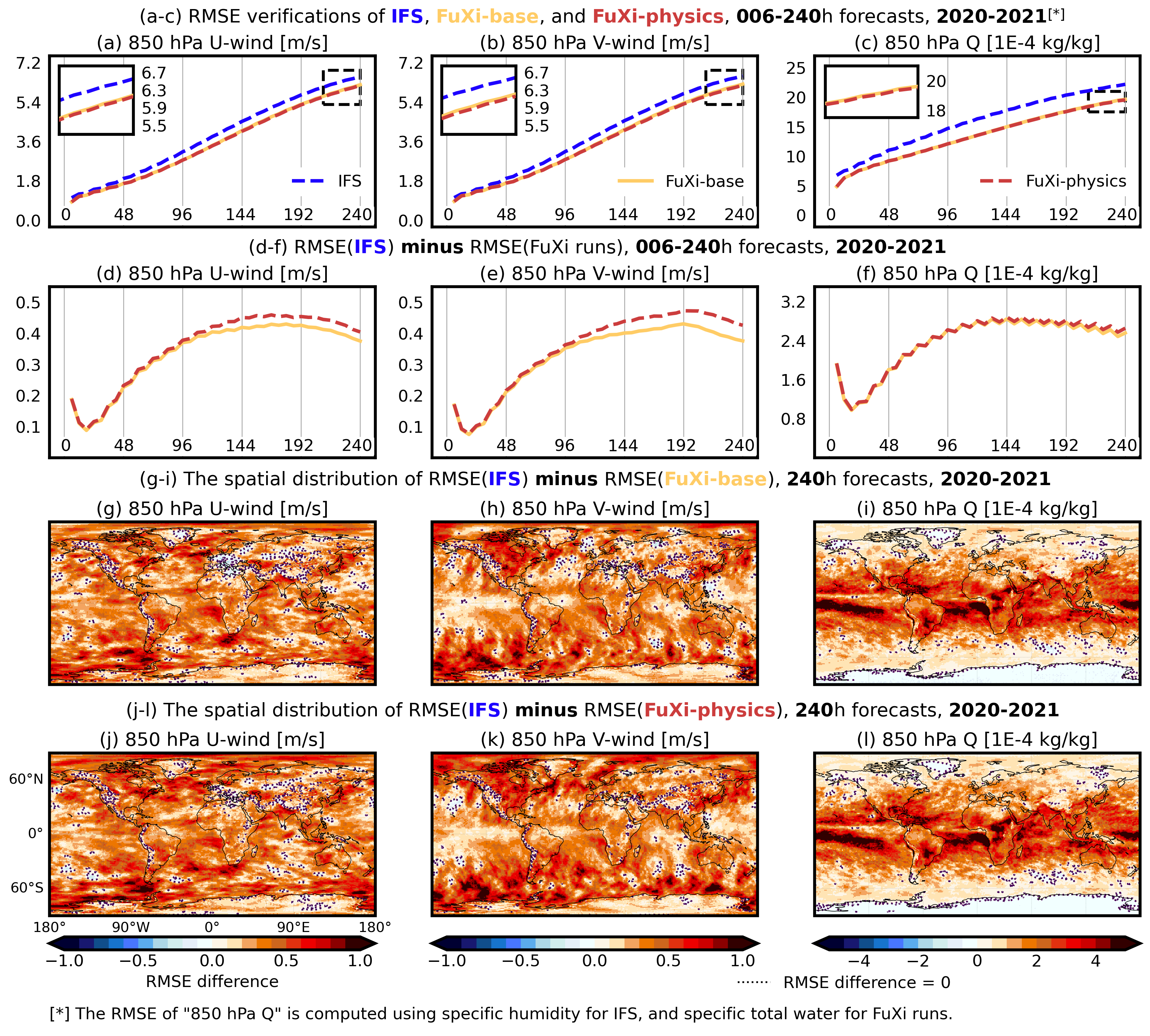}
    \caption{As in Figure~\ref{fig6}, but for 850 hPa zonal wind (``U-wind''), meridional wind (``V-wind''), and specific total water (``Q''). For IFS-HRES, the RMSE of Q is calculated using specific humidity. Note that panels (c), (f), (i), and (j) have scales of $10^{-4}$ kg/kg.}
    \label{fig7}
\end{figure}

Figure~\ref{fig6} and Figure~\ref{fig7} provide RMSE verifications for mean sea level pressure and upper air variables on 500 and 850 hPa levels. The two FuXi runs showed good performance. Their RMSE curves generally track IFS-HRES but with lower RMSE values (Figure~\ref{fig6}a-c and Figure~\ref{fig6}a-c). The RMSE differences between IFS-HRES and FuXi runs are examined, with positive and higher values pointing to performance gains. For all verified variables, their RMSE differences are generally positive. They typically show a small decrease from 6 to 24 hours, then increase rapidly from 48 to 192 hours, and reach a plateau for longer forecast lead times. The decrease of RMSE differences in 6-24 forecasts is likely explained by the multistep training procedures, which tend to improve the stability and long-term performance of AIWP models by compensating for their short-term forecast skill \cite{schreck2024community}. For 500 hPa air temperature, this issue has caused some small negative RMSE differences on 18 and 24-hour forecasts (i.e., IFS-HRES are slightly better in these two forecast lead times), but for other forecast lead times and variables, the two FuXi runs are verified to be more skillful than IFS-HRES (Figure~\ref{fig6}d-f and Figure~\ref{fig7}d-f). At 240 hours, the two FuXi runs provide roughly a 10\% performance increase for 850 hPa specific total water (Figure~\ref{fig7}f) and a 5-8\% performance increase for other variables.

The spatial distributions of RMSE differences are also verified; again, positive RMSE differences indicate a performance increase of the two FuXi runs over IFS-HRES. For 850 hPa specific total water, its performance increase is centered in the trade wind belt and the Inter Tropical Convergence Zone (ITCZ), where a high concentration of atmospheric moisture exists (Figure~\ref{fig7}i). For 500 hPa geopotential height, air temperature and mean sea level pressure, their performance increase is located at polar regions, notably the South Pole and the Southern Ocean (Figure~\ref{fig6}h and i). For 850 hPa zonal and meridional winds, the spatial distribution of their RMSE performance increase is generally even (Figure~\ref{fig6}g; Figure~\ref{fig7}g and h), and with the RMSE performance gain of the two FuXi runs being mostly positive. Some regions have reported negative performance relative to IFS-HRES. For 500 hPa geopotential height, air temperature, and mean sea level pressure, the two FuXi runs are suboptimal in the North Atlantic, Europe, and Siberia. For 850 hPa zonal and meridional winds, several negative performance zones are collocated with large-scale terrain features, such as the Andes and Tibet Plateau. These areas are highlighted using dotted lines and blue colored shades in Figure~\ref{fig6}g-l and Figure~\ref{fig7}g-l.

The conservation schemes are found to play a positive role in the verification of mean sea level pressure and upper-air variables. Compared to FuXi-base, FuXi-physics presented more performance gains over the IFS-HRES with larger RMSE differences. These performance gains are evident after day 4, and they tend to increase with forecast lead times (Figure~\ref{fig6}d-f and Figure~\ref{fig7}d-f). For globally averaged RMSE scores, 500 hPa geopotential height and air temperature received the largest performance increase. The spatial distribution of day 10 RMSE difference revealed that certain mid- and high-latitude regions have received larger benefits. For 500 hPa air temperature, geopotential height, and mean sea level pressure, FuXi-physics largely improved the RMSE scores in Europe, including most areas on the western side of the Black Sea and the Scandinavian (c.f., the blue color shade with dots in Figure~\ref{fig6}g-i and j-l). For 850 hPa zonal and meridional winds, FuXi-physics also showed better performance for areas around the Mediterranean Sea (c.f., Figure~\ref{fig7}g-i and j-l). For 850 hPa specific total water, FuXi-physics improved over FuXi-base slightly for day 6 and onward, but the improvement does not have a specific spatial pattern. We found that many tropical grid cells have received a small and evenly distributed performance increase.

\begin{figure}
    \centering
    \includegraphics[width=\columnwidth]{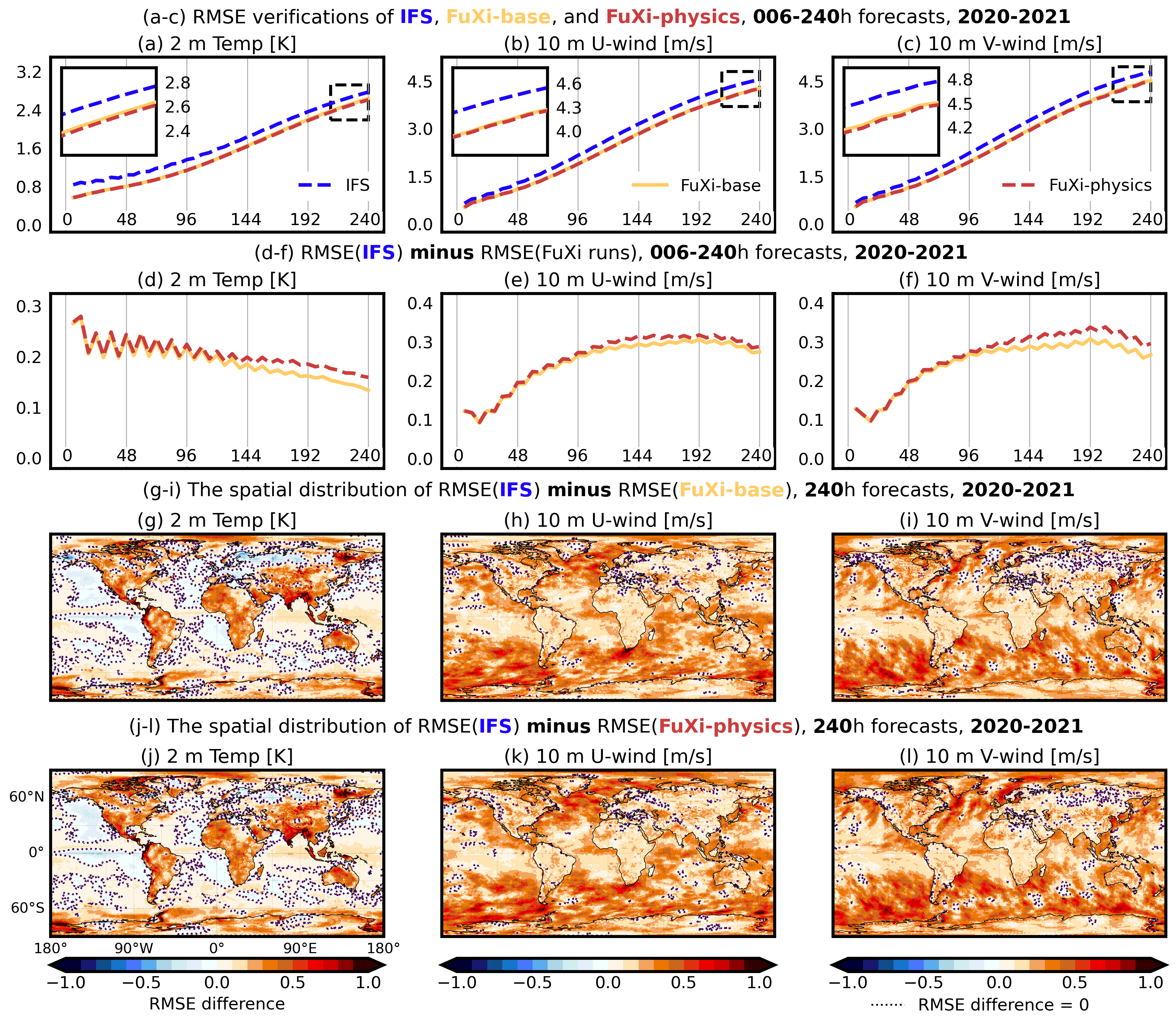}
    \caption{As in Figure~\ref{fig3}, but for 2-m air temperature (``2-m Temp''), 10-m zonal wind (``10-m U-wind''), and 10-m meridional wind (``10-m V-wind'').}
    \label{fig8}
\end{figure}

Figure~\ref{fig8} provides verification results for near-surface prognostic variables: 2-m air temperature, 10-m zonal wind, and 10-m meridional wind. The two FuXi runs showed better domain-averaged RMSE performance than IFS-HRES (Figure~\ref{fig8}a-c). For 10-m winds, the outperformance increases with forecast lead times but with a small decrease for 6-24 hour forecasts, similar to that of upper-air variables (Figure~\ref{fig8}e-f). The spatial distributions of the 10-m winds performance increase are comparable to that of 850 hPa winds, with the areas around the Mediterranean Sea receiving benefits (Figure~\ref{fig8}h-i and k-l). For 2-m air temperature, the performance increase of the two FuXi runs is stationary or slightly decreased at longer forecast lead times, but overall, it is clearly better than that of the IFS-HRES (Figure~\ref{fig8}d). The spatial distribution of the RMSE difference shows that the 2-m air temperature performance gains are evident only in land grid cells. The performance of the two FuXi runs is on the same level as IFS-HRES for mid-latitude oceans and slightly worse for tropical oceans (Figure~\ref{fig8}g and j).

FuXi-physics outperformed FuXi-base, with larger performance gains found at longer forecast lead times. For 2-m air temperature, FuXi-physics improved performance in the North Hemisphere high latitude areas, including Scandinavia and Siberia. For 10-m zonal wind, FuXi-physics does not provide a clear spatial signal for its performance gains, but its verification contains fewer negative RMSE patterns on various land grid cells. For 10-m meridional wind, some improvements can be found along the Alaska Coast and Iceland.

In summary, for the key pressure level and near-surface variables, both FuXi-base and FuXi-physics outperformed the IFS-HRES, with performance gains generally increasing with forecast lead time. The results align with other AIWP model studies \cite<e.g.>{rasp2024weatherbench,bi2023accurate,schreck2024community} indicating that the FuXi architecture and training have been conducted successfully. Taking FuXi-base as a reference, FuXi-physics, the configuration equipped with the conservation schemes, provides additional performance gains, typically starting from day 4. For many variables, the contribution of conservation schemes is not a unified forecast skill increase but has its own spatial signals, with high-latitude and land grid cells receiving larger benefits. 

\subsubsection{Flux form variables}\label{sec422}

\begin{figure}
    \centering
    \includegraphics[width=\columnwidth]{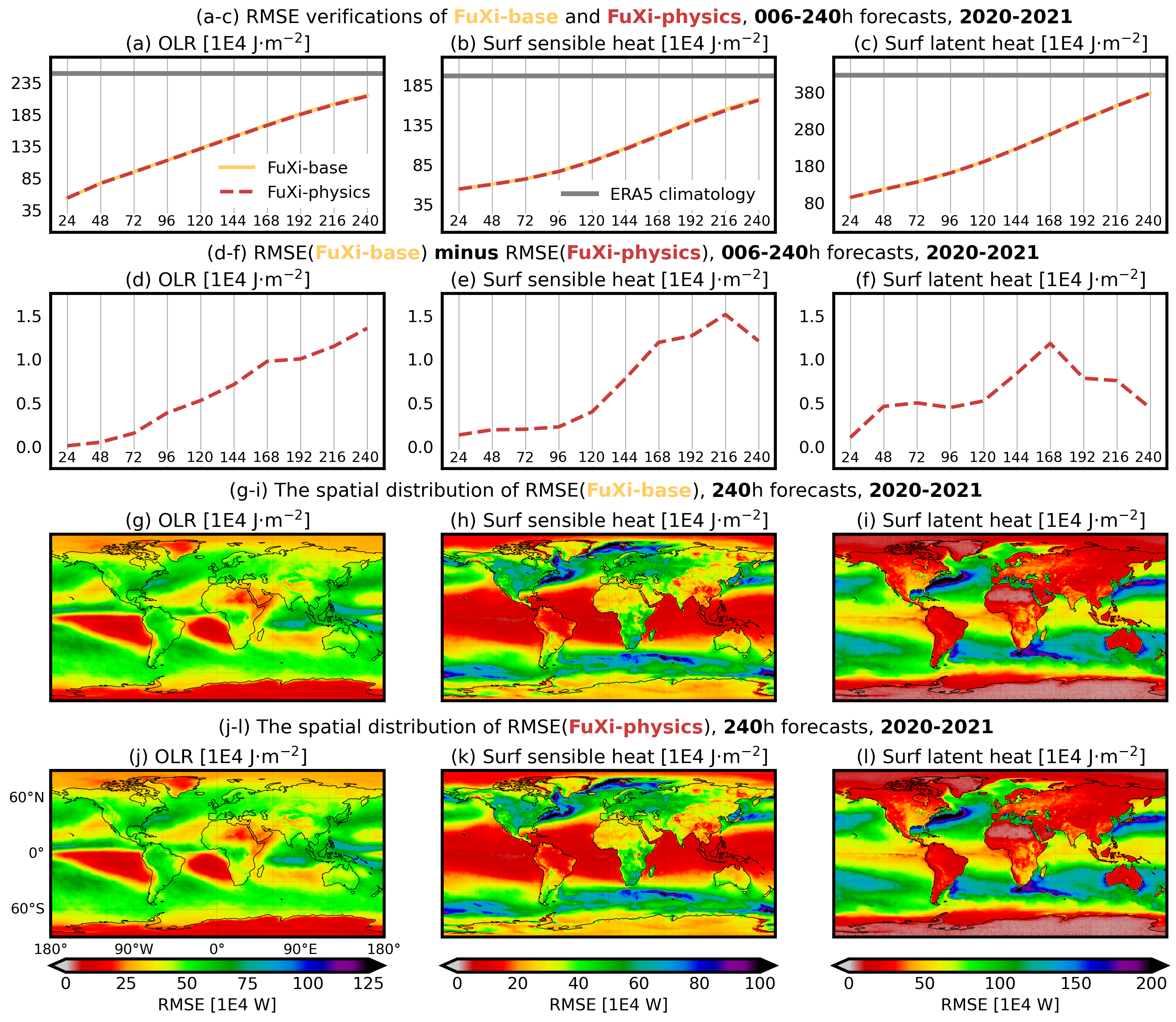}
    \caption{RMSE verifications of FuXi-base (orange solid line) and FuXi-physics (red dashed line) for all 00 and 12Z initializations in 2020-2021. (a-c) Domain-weighted averaged RMSE by forecast lead times for outgoing longwave radiation (``OLR''), surface net sensible heat fluxes, and surface net latent heat fluxes, respectively. Lower is better. Gray solid is RMSE computed from the ERA5 climatology. (d-f) The RMSE of FuXi-base minus FuXi-physics. Positive and higher means FuXi-physics is better, vice versa. (g-i) The spatial distribution of RMSE of FuXi-base on day 10 forecasts. (j-l) as in (g-i), but for FuXi-physics. All panels have scales of $10^4$ $\mathrm{J\cdot m^{-2}}$.}
    \label{fig9}
\end{figure}

Flux form variables describe the accumulated amounts of mass and energy that enter or leave the atmosphere. In this study, they are modeled as diagnostic variables. Figure~\ref{fig9} provides the deterministic verification results of three key flux form variables: outgoing longwave radiation at the top of the atmosphere and the sensible and latent heat fluxes at the surface. 

The two FuXi runs modeled flux form variables successfully. The domain-averaged RMSE curves show a steady and linear increase with forecast lead time. The trends are similar to other prognostic variables verified in Section \ref{sec421}. The RMSE of the two FuXi runs is lower than climatology scores, indicating more skillful forecasts than climatology references. The difference between the two FuXi runs is minimal. FuXi-physics performed slightly better than FuXi-base in terms of domain-averaged RMSE, but the amount of performance gains is only 0.3-0.5\%. The two FuXi runs also share the same spatial distributions of RMSE. For outgoing longwave radiation, the highest RMSE is found over the western tropical Pacific Ocean. This is explained by the strong and frequent convective activities in this area. For sensible and latent heat fluxes, their highest RMSE is found in the mid- and high-latitude ocean, notably the North Atlantic Ocean and Southern Oceans. This is explained by the global energy transport of ocean currents from tropical to high-latitude regions. That said, the spatial distributions of RMSE values produced by the two FuXi runs are not impacted by unphysical artifacts and noise. Regions having higher energy fluxes also have higher RMSE values.

Overall, the two FuXi runs produced healthy RMSE curves and spatial distributions, and the performance is better than the climatology reference on day 10. FuXi-physics does not outperform FuXi-base, indicating that the conservation schemes may not have a clear effect on diagnostic variables that are not correct directly (i.e., total precipitation is an exception because it is corrected directly to close the moisture budget conservation residuals).

\subsubsection{Zonal energy spectrum}\label{sec423}

\begin{figure}
    \centering
    \includegraphics[width=\columnwidth]{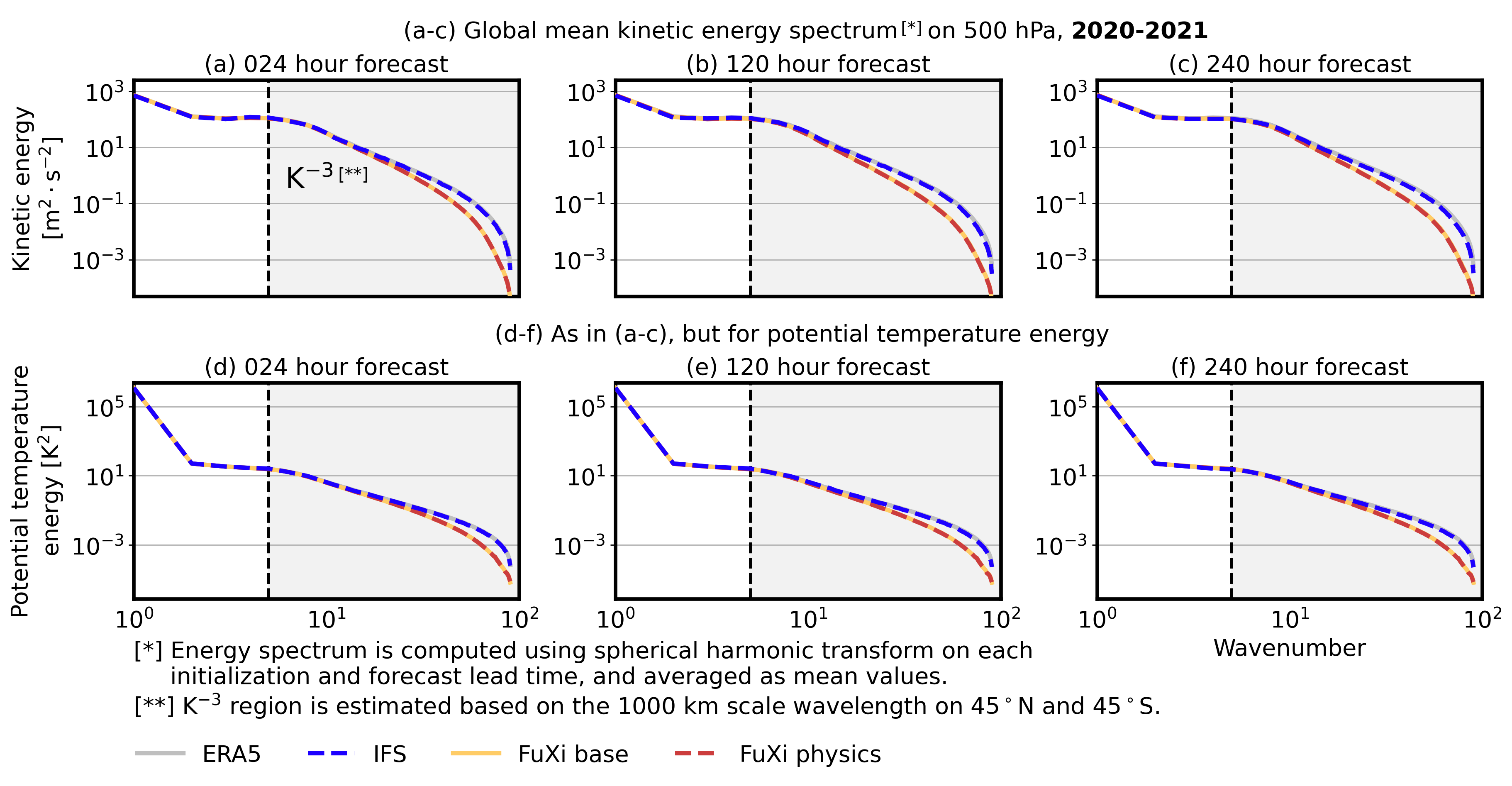}
    \caption{Verifications of 500 hPa global mean kinetic energy and potential temperature energy spectra in the ERA5 (gray solid line), IFS-HRES (blue dashed line), FuXi-base (orange solid line), and FuXi-physics (red dashed lines). For forecasts, the result is averaged from all 00 and 12Z initializations in 2020-2021.}
    \label{fig10}
\end{figure}

Global 500 hPa kinetic energy and potential temperature energy are verified in Fig~\ref{fig10} as the power spectrum varies by zonal wavenumbers. For forecast lead times of 24, 120, and 240 hours, the zonal energy spectrum of IFS-HRES closely resamples that of the ERA5 dataset, showing a steady energy dissipation in the synoptic scale. For the two FuXi runs, their zonal energy spectrum diminished faster than IFS-HRES for wavenumber 40 and beyond, and the amount of dissipated energy increases with forecast lead time on the same wavenumber. This problem is expected and has been reported in \citeA{rasp2024weatherbench} and \citeA{schreck2024community}. AIWP models tend to produce spatially smoothed forecasts, and the level of smoothness is higher for longer forecast lead times. This behavior would lead to the loss of spectral resolution and the ability to capture smaller-scale features. 

FuXi-physics and FuXi-base share the same problems in Fig~\ref{fig10}, indicating that the use of conservation schemes cannot preserve the loss of fine-scale features. On the positive side, this also means the conservation schemes will not make the situation worse. That said, the conservation schemes will not introduce additional blurry effects, and the RMSE improvements of prognostic variables in FuXi-physics are actual forecast skill gains without compensating on the blurriness of spatial patterns.

\section{Discussion}\label{sec5}

\subsection{What can conservation schemes do and why?}\label{sec51}

AIWP models have shown great success in medium-range weather forecasts. In this study, the representative of these pure AIWP models, FuXi-base, achieved good deterministic verification scores, typically outperforming the IFS-HRES, and aligned with other AIWP models on Weatherbench2. Even for the verification of global mass and energy conservation, its residual terms do not go out of bounds exponentially. Taking a competitive pure AIWP model, FuXi-base, as a reference, FuXi-physics is verified to be better in many aspects. In this section, several key verification highlights and explanations are addressed. The discussion here is also responses to the research questions of this study.

\subsubsection{Mass and energy conservation improvements}\label{sec511}

The most important thing that conservation schemes can do is ensure that the internal state of AIWP models does not drift in unphysical ways over time. FuXi-physics produced clearly smaller conservation residuals, and its key conservation properties are more stable during iterative forecasts. The correct conservation behavior itself does not translate to a significant forecast skill increase (although it does improve the forecast skills), but it lays a foundation for the incorporation of other schemes relying on atmospheric science domain knowledge, including the spectral stochastic kinetic energy backscatter scheme in \citeA{berner2009spectral}, and the moist dynamics in \citeA{gutmann2016intermediate}. Many modeling schemes are simple enough to be re-phased for neural networks and contribute to AIWP, but they also lack a full dynamical solution; a direct integration of them can cause stability issues. The conservation schemes of this study are hard physical constraints that regulate the behavior of AIWP models. They force the AIWP model to conserve global mass and energy, and thus, draw a bottom line for AIWP models to avoid certain unphysical/unstable output patterns. This potentially creates a space for future studies to incorporate a wide range of advanced physics-based schemes in AIWP models. In addition, this benefit is not limited to medium-range weather forecasts. Conservation behaviors are critical measures in climate model projections. We believe the conservation schemes of this study can also provide solid support to the development of AI-based climate models.

\subsubsection{Total precipitation improvements}\label{sec512}

Another important achievement of the conservation schemes is the improvement of total precipitation forecasts. FuXi-physics achieved clearly better SEEPS verification scores compared to FuXi-base, and the main reason for this performance gain is fully explainable---a verified reduction of drizzle bias (see Section \ref{sec424}). 

Many AIWP models have blurriness issues. For total precipitation, it causes the overestimation of drizzle \cite<e.g.>{rasp2024weatherbench}. The conservation schemes can tackle this problem because they are operated based on the global sum rather than gridpoint-wise quantities. In the moisture budget conservation scheme, the global sum of total precipitation must match with other conservation properties. If the AIWP model overestimates drizzle and underestimates dry areas, its global sum of total precipitation will violate the conservation laws, and corrections will be applied--this correction is assigned to close the conservation budget only, and it almost surely will increase the mean squared error training loss--so the AIWP model will be penalized by its drizzle bias. 

The verification target of total precipitation is ERA5, which is known for overestimating drizzle \cite<e.g.>{hersbach2020era5,wu2022validation}. Here, the identified benefit of conservation schemes is the reduction of drizzle bias. It is an aspect of improvement that plays against, rather than overfitting to, the caveats of ERA5. Thus, this benefit is likely independent of the use of ERA5 as training and verification targets. Moreover, implementing the conservation schemes of this study can be an effective way for AIWP models to battle the drizzle bias introduced by the ERA5 training data.

\subsubsection{Prognostic variable improvements}\label{sec513}

\begin{figure}
    \centering
    \includegraphics[width=\columnwidth]{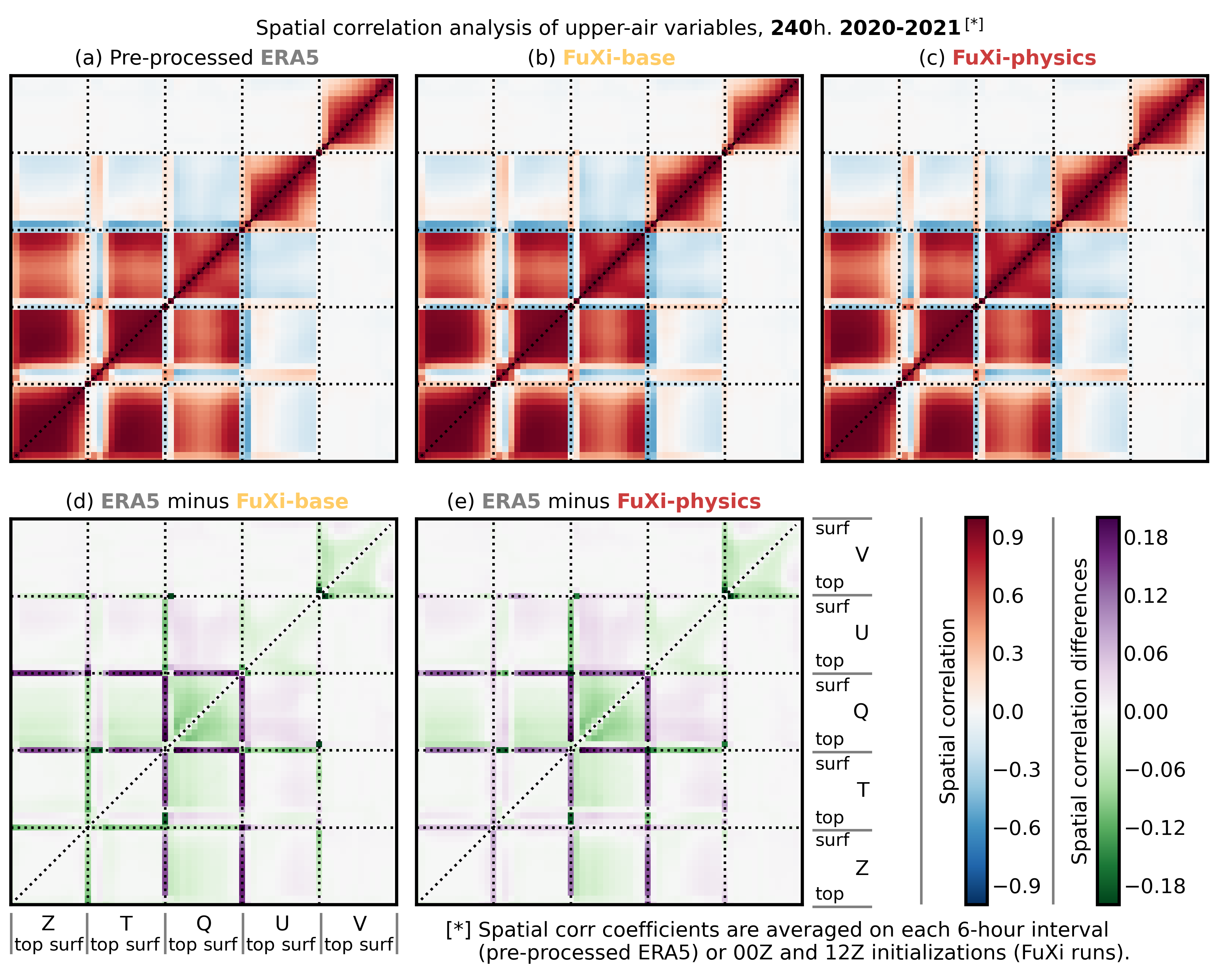}
    \caption{Spatial correlation analysis on pressure level geopotential height (``Z''), air temperature (``T''), specific total water (``Q''), zonal wind (``U''), and meridional wind (``V''); from 1 hPa level (``top'') to 1000 hPa level (``surf''). (a) Self- and cross-correlation of ERA5 variables in 2020-2021. (b) As in (a), but for FuXi-base on 240-hour forecasts. (c) As in (b), but for FuXi-physics. (d) The difference of correlation coefficients between ERA5 and the 240-hour forecast of FuXi-base. (e) As in (d), but for FuXi-physics.}
    \label{fig12}
\end{figure}

The conservation schemes have also been found to improve the performance of a wide range of prognostic variables, notably 500 hPa geopotential height and mean sea level pressure. These variables are not corrected by the conservation schemes directly, but they are still verified to be better in FuXi-physics, and the performance gains are more evident at longer forecast lead times. 

The integration of physical constraints in AIWP models is in its early stage, so most of these performance increases do not have established explanations. First of all, this calls for the importance of explainable AI studies in the future, and here, we will discuss some possible reasons for these performance gains. 

The conservation schemes are verified to suppress the oscillation of mass and energy tendency terms and correct them to a zero-centered position (see Section \ref{sec41}). This means that the conservation schemes can prevent the atmosphere from losing humidity and temperature during the iterative forecasts of AIWP models. This benefit is accumulative. When the iterative forecast step increases, the gap between a pure AIWP model and the ones with conservation schemes will increase, and thus, more performance gains can be found at longer forecast lead times.  

Another possible reason for the performance gains is the cross-correlation of prognostic variables. Prognostic upper-air variables have spatial correlations, and AIWP models are capable of fitting these correlation patterns. This is evident in Figure~\ref{fig12} where the two FuXi runs maintained the self- and cross-correlations of upper-air variables as in the pre-processed 1.0$^\circ$ ERA5 except the 1 hPa atmospheric top [technical details of spatial correlation refers to \citeA{schreck2024community}]. With better air temperature and humidity forecasts introduced by the conservation schemes, their impact can be propagated to other variables, such as geopotential height, through the learned cross-variable relationships.


\subsection{How can existing AIWP models implement conservation schemes?}\label{sec52}

The conservation schemes of this study are designed as separated modules, and they can be applied to any state-of-the-art AIWP model (Figure~\ref{fig2}a). Section \ref{sec51} has summarized the major benefits of having the conservation schemes. FuXi is used as an example, but again, these benefits are not tied to a specific architecture.

The conservation schemes are not post-processing operations. When an AIWP model is equipped with the conservation schemes, it needs to be re-trained. Ideally, the re-training shall start from scratch, but fine-tuning could be a situational alternative. Part of the reasons are provided in Section \ref{sec512}--the conservation schemes can improve the model training by penalizing the drizzle bias--and there could be other positive effects that we have not discovered.

The conservation schemes need to be implemented in the specified order (see Section \ref{sec31}), but part of them can be discarded. For example, the global total atmospheric energy conservation scheme can be ignored if the training dataset does not provide all required energy fluxes (see Table \ref{tab1}). Since most of the AIWP models are trained using the ERA5, the conservation schemes of this study are easy to implement.

\subsection{What are the remaining issues?}\label{sec53}

The conservation schemes cannot improve the spectral resolution of kinetic energy and potential temperature energy (see Section \ref{sec423}). This is because closing the global energy conservation budget does not guarantee better energy dissipation across scales for AIWP models. At the same time, however, the conservation schemes would not make this problem worse, as the FuXi-physics is verified to have the same zonal energy spectrum as that of the FuXi-base. This means that conservation schemes can collaborate with other solutions to tackle this problem. With the conservation schemes turned on, many physics-based parameterizations can be applied to estimate the contribution of atmospheric energy from subgrid-scale processes without taking the risk of violating global energy conservation. From this perspective, the conservation schemes of this study, although not helping by themselves, can still be a critical step for solving the spectral fidelity problem. 

Most flux form variables do not receive large performance gains from the conservation schemes. The exception is total precipitation, which is corrected directly for the conservation of global moisture budget. A possible explanation of this issue is the cross-variable correlation, which has possibly contributed to the performance increase of prognostic variables (see Section \ref{sec512}). Flux form variables are modeled as diagnostics, and they likely exhibit fewer correlations with humidity and temperature. Thus, with the conservation schemes contributing to prognostic variables, this benefit may not propagate to flux form variables. We think this issue is not really a problem. Although the performance gain is relatively small, FuXi-physics is still verified to be better for flux form variables. That said, it is not a performance downgrade. Many flux form variables are crucial inputs for the coupling between the atmospheric model and other earth system components. Once the AI-based atmospheric modeling is advanced to the coupled simulation stage, more avenues will be opened for improving the performance of flux form variables.


\subsection{Other lessons learned}

A lesson learned from this study is that physical relationships, such as global mass and energy conservation, can be specified outside of the AIWP model structure but still provide positive impacts on the model. The learning process of AIWP models is generally a  ``black box''. Having the key physical relationships deployed outside the box can improve the transparency and trustworthiness of AIWP models. We believe this is also a way to bring more domain knowledge to AIWP models. The numerical modeling communities have established much knowledge to represent the physical processes of the atmosphere. Transferring this valuable knowledge from NWP to AIWP models can be a direction of improvements in the future.   

Three variables are corrected in the conservation schemes: specific total water, total precipitation, and air temperature. They are chosen in this study for specific reasons. For global dry air mass conservation, specific total water is corrected because it is the root cause of the conservation residuals. AIWP models produce negative moisture without considering its column-integrated properties, which leads to the violation of global mass conservation. Correcting specific total water targets the problem directly. For the conservation of global moisture budget, correcting total precipitation and evaporation are equally effective. However, total precipitation is more sensitive to the multiplicative correction ratio because its spatial distribution is highly inhomogeneous and positively skewed. Correcting total precipitation will give the AIWP model larger penalties for violating the conservation law, and thus, improve their training performance. For global energy conservation, air temperature is corrected. This is because thermal energy has the first-order importance in total atmospheric energy. In the early stage of model training, thermal energy is the only term that can close the energy budget. When the training losses of all variables are high, exhausting latent heat energy or other components completely may not be sufficient to maintain the conservation. Thus, air temperature is chosen because it controls the total amount of thermal energy in the atmosphere. The above consideration shows that when transferring physics-based schemes from numerical systems to AIWP models, adjustments are needed based on the behavior of neural networks. 

Another lesson learned is that the RMSE improvement of AIWP models relative to IFS-HRES is region-dependent. This has been identified in Section \ref{sec421}. Most of the current AIWP model studies are focused on the global-averaged performance change. However, we think it is equally important to provide a clear picture of how the performance of AIWP models varies per area. This has motivated us to verify the spatial distribution of RMSE differences. An interesting finding is that the two FuXi runs perform better for 2-m air temperature on land than over the ocean (Figure~\ref{fig9}g and j). Other AIWP models may have their own spatial distribution of errors, but for the two FuXi runs here, incorporating sea surface temperature as forcing can be a solution.          

Finally, as shown in Section \ref{sec41} and many other studies \cite<e.g.>{trenberth2018applications,mayer2021consistency}, ERA5 does not conserve global mass and energy strictly. Pure AIWP models, such as FuXi-base, will inherit the magnitude of conservation residuals as presented in the ERA5, and worse---the averaged moisture and energy tendency terms of the ERA5 are generally zero-centered, but for FuXi-base, they are not. We believe ERA5 is the best training data for developing AIWP models, especially with the conservation schemes as a fix. Still, process-based studies should be conducted in the future to examine how the caveats of ERA5 would impact AIWP models. This can potentially be done by introducing other training data for comparisons.

\section{Conclusion}\label{sec6}

A set of novel physics-based schemes are proposed for AIWP models to guide them in producing forecasts that can conserve global dry air mass, moisture budget, and total atmospheric energy. The conservation schemes are designed on a modular basis, and they can be applied to AIWP models regardless of their architecture. 

To evaluate the effectiveness of these conservation schemes, numerical experiments are conducted using FuXi, an AIWP model, as an example. Two configurations are prepared: FuXi-base, maintaining its original design, and FuXi-physics, equipped with the conservation schemes. The two FuXi configurations are trained using the pre-processed 1.0$^\circ$ ERA5 and verified in 2020-2021. Verification results demonstrate that the conservation schemes are successfully implemented. FuXi-physics exhibited largely reduced conservation residuals, as well as more stable and zero-centered moisture and energy tendency terms.

Deterministic verifications are also conducted, and it shows that the FuXi-physics outperformed the FuXi-base, owing to the implementation of conservation schemes. The biggest performance gain is total precipitation (5-7\% forecast skill increase), verified using SEEPS. The FuXi-physics are found to have largely reduced drizzle bias and better separations of dry and light precipitation areas. These improvements are attributed to the global moisture budget conservation scheme, which penalizes the overestimation of drizzle. 

Other prognostic variables, including 500 hPa geopotential height, 500 hPa air temperature, 850 hPa specific total water, and mean sea level pressure, also benefited from the conservation schemes, with typical performance gains of 5\%, and up to 10\% for specific total water. The forecast skill benefits are more evident at longer forecast lead times. This is mainly explained by the ability of conservation schemes to prevent the model from progressively losing moisture and energy during iterative forecasting. Additionally, the performance gains likely propagate through cross-variable correlations learned by the AIWP model.

The conservation schemes cannot improve the spectral resolution of kinetic energy and potential temperature energy. However, they also do not introduce additional blurriness to the forecasts. With the conservation schemes helping AIWP models avoid certain unphysical/unstable forecast states, other methods can be applied to solve the energy spectrum problem without taking the risk of violating conservation properties.

In conclusion, the conservation schemes presented in this study demonstrate how atmospheric science domain knowledge can be effectively integrated into AIWP models. The schemes can adjust the global mass and energy conservation behaviors of the AIWP models and are verified to bring performance gains in both the stability of conservation properties and the deterministic verification scores. Taking the conservation schemes here as a foundation, more specialized, physics-based schemes can be introduced to AIWP models to tackle challenges in medium-range weather forecasting and beyond. 

\appendix
\section{The conservation schemes}\label{A1}

The conservation schemes of this study are designed to correct AIWP model outputs and enforce the conservation of global dry air mass, moisture budget, and total atmospheric energy. In this section, the technical details of these conservation schemes are summarized.

\subsection{Preliminaries}\label{A11}

For a given quantity $X(\phi, \lambda)$ that varies with latitude ($\phi$) and longitude ($\lambda$), its global weighted sum $\overline{X}$ is defined as:

\begin{equation}\label{sec2_eq1}
\overline{X} = \int_{0}^{2\pi} \int_{-\pi/2}^{\pi/2}  X \cdot R^2 \cdot d\left(\sin \phi \right) d\lambda
\end{equation}

\noindent
Where $R$ is the radius of earth. For gridded data, equation (\ref{sec2_eq1}) can be converted to a discrete form:

\begin{equation}\label{sec2_eq2}
\overline{X} = \sum_{i_\phi=0}^{N_\phi} \sum_{i_\lambda=0}^{N_\lambda} {\left[X \cdot R^2 \cdot \Delta\left(\sin \phi \right) \cdot \Delta \lambda\right]}_{i_\phi,i_\lambda}
\end{equation}

\noindent
Where $i_\phi=\left\{0, 1, \ldots, N_\phi\right\}$ and $i_\lambda=\left\{0, 1, \ldots, N_\lambda\right\}$ are indices of latitude and longitude grids, respectively. 

The terms $\Delta\left(\sin \phi \right)$ and $\Delta \lambda$ are computed using the second-order difference for central grid cells and the forward difference for edge grid cells. Hereafter, the global weighted sum is denoted as $\overline{X} = \text{SUM}\left(X\right)$ 

For a quantity $X(z)$ that varies with height $z$, its mass-weighted vertical integral can be converted to a pressure level integral using the hydrostatic equation:

\begin{equation}\label{sec2_eq3}
\int_{0}^{\infty}{\rho X}dz = \frac{1}{g}\int_{p_s}^{0}Xdp \approx \frac{1}{g}\int_{p_1}^{p_0}Xdp
\end{equation}

\noindent
Where $g$ is gravity, $p_s$ is surface pressure, $p_1$ and $p_0$ are the highest and lowest discrete pressure levels available in a dataset, respectively.

\subsection{Global dry air mass conservation scheme}\label{A12}

For a given air column, the tendency of its dry air mass is determined by the divergence of the vertically integrated dry air mass flux:

\begin{equation}\label{sec2_eq4}
\frac{1}{g}\frac{\partial}{\partial t}\int_{p_1}^{p_0}{\left(1-q\right)}dp = -\mathbf{\nabla} \cdot \frac{1}{g} \int_{p_1}^{p_0}{\left[\left(1-q\right)\mathbf{v}\right]}dp
\end{equation}

\noindent
Where $\mathbf{v}$ is the velocity, and $q$ is the total atmospheric moisture, approximated using specific total water (see Table \ref{tab1}).

For global sum, the divergence term in equation (\ref{sec2_eq4}) is zero for incompressible atmosphere. Thus, the total amount of global dry air mass ($\overline{M_d}$) is conserved regardless of time:

\begin{equation}\label{sec2_eq5}
\frac{\partial}{\partial t}\overline{M_d} = \frac{\partial}{\partial t}\text{SUM}\left[\frac{1}{g}\int_{p_1}^{p_0}{\left(1-q\right)}dp\right] = \epsilon_d
\end{equation}

\noindent
Where $\epsilon_d$ is the residual term that violates the global dry air mass conservation.

Given two forecast steps $\Delta t = t_1 - t_0$ with $t_0$ representing the analyzed initial condition and $t_1$ representing an arbitrary validation time, equation (\ref{sec2_eq5}) can be written as:

\begin{equation}\label{sec2_eq6}
\overline{M_d\left(t_0\right)} - \overline{M_d\left(t_1\right)} = \epsilon_d
\end{equation}

\noindent
During the correction stage, $q$ can be adjusted to ensure $\epsilon_d=0$ using a multiplicative ratio:

\begin{equation}\label{sec2_eq7}
q^*\left(t_1\right) = 1 - \left[1 - q\left(t_1\right)\right] \frac{\overline{M_d\left(t_0\right)}}{\overline{M_d\left(t_1\right)}}
\end{equation}

\noindent
Where $q^*\left(t_1\right)$ is the corrected $q$ on $t_1$. The same multiplicative correction is applied to $q$ on all grid cells. 

An improved version of equation (\ref{sec2_eq7}) is to correct $q$ below a level threshold $L_0$ (e.g., 600 hPa level) only, leaving $q$ values above $L_0$ unchanged. This modified correction is given by:

\begin{equation}\label{sec2_eq72}
q^*\left(t_1,L_{0{\downarrow}}\right) = 1-\left[1 - q\left(t_1,L_{0{\downarrow}}\right))\right]\frac{\overline{M_d\left(t_0\right)} - \overline{M_d\left(t_1,L_{0{\uparrow}}\right)}}{\overline{M_d\left(t_1,L_{0{\downarrow}}\right)}}
\end{equation}

\noindent
Where $L_{0{\downarrow}}$ and $L_{0{\uparrow}}$ denote the regions below and above $L_0$, respectively. In equation (\ref{sec2_eq72}), $q\left(t_1,L_{0{\downarrow}}\right)$ is corrected, while $q\left(t_1,L_{0{\uparrow}}\right)$ remains unaffected.

\subsection{Global moisture budget conservation scheme}\label{A13}

For a given air column, the tendency of its total precipitable water ($M_v$) is determined by the divergence of vertically integrated moisture flux, total precipitation, and evaporation:

\begin{equation}\label{sec2_eq8}
\frac{\partial}{\partial t}M_v = \frac{1}{g}\frac{\partial}{\partial t}\int_{p_1}^{p_0}{q}dp = -\mathbf{\nabla} \cdot \frac{1}{g} \int_{p_1}^{p_0}{\left(\mathbf{v}q\right)}dp - E - P
\end{equation}

\noindent
Where $q$ is approximated using specific total water, $E$ and $P$ are evaporation and total precipitation with units of $\mathrm{kg\cdot m^{-2} \cdot s^{-1}}$, respectively.

For global sum, the divergence term in equation (\ref{sec2_eq8}) is zero, and the global sum of $M_v$ is balanced by $\overline{E}$ and $\overline{P}$, subject to a residual term $\epsilon_m$ that violates the conservation:

\begin{equation}\label{sec2_eq9}
-\overline{\left(\frac{\partial M_v}{\partial t}\right)} - \overline{E} - \overline{P} = \epsilon_m
\end{equation}

\noindent
Note that $\overline{P}$ is always positive for downward precipitation. $\overline{E}$ is mostly negative for upward evaporation.

During the correction stage, $P$ is adjusted to ensure $\epsilon_m=0$ using a multiplicative ratio:

\begin{equation}\label{sec2_eq10}
P^*\left(t_1\right) = P\left(t_1\right)\frac{\overline{P^*\left(t_1\right)}}{\overline{P\left(t_1\right)}}, \quad\overline{P^*\left(t_1\right)} = -\overline{\left[\frac{M_v\left(t_1\right) - M_v\left(t_0\right)}{\Delta t}\right]} - \overline{E\left(t_1\right)}
\end{equation}

\noindent
Where $\overline{P^*\left(t_1\right)}$ is the corrected global sum of total precipitation required to close the moisture budget. The same multiplicative ratio is applied to all grid cells.

\subsection{Global total atmospheric energy conservation scheme}\label{A14}

For a given air column, its vertically integrated total atmospheric energy ($A$) is defined as follows:

\begin{equation}\label{sec2_eq11}
A = \frac{1}{g}\int_{p_1}^{p_0}{\left(C_pT+L_v q+\Phi_s+k\right)}dp
\end{equation}

\noindent
The terms on the right side of the equation (\ref{sec2_eq11}) are thermal energy, latent heat energy, potential energy, and kinetic energy, respectively. $L_v$ is the latent heat of vaporization, and $\Phi_s$ is the geopotential at the surface. Kinetic energy ($k$) is defined as $k=0.5\left(\mathbf{v} \cdot \mathbf{v}\right)$. The specific heat capacity of air at constant pressure ($C_p$) is defined as $C_p=C_{pd}(1-q)+C_{pv}q$. The formulation of equation (\ref{sec2_eq11}) has some limitations, which will be addressed in a separated section below.

The tendency of $A$ is determined by the divergence of vertically integrated moist static energy ($h=C_pT+L_v q+\Phi$), kinetic energy, and other energy sources and sinks:

\begin{equation}\label{sec2_eq12}
\frac{\partial}{\partial t}A = -\mathbf{\nabla}\cdot\frac{1}{g}\int_{p_1}^{p_0}{\mathbf{v}\left(h+k\right)}dp = R_T - F_S
\end{equation}

\noindent
Where $R_T$ and $F_S$ are net radiation and energy fluxes on the top of the atmosphere and the surface.

\begin{equation}\label{sec2_eq13}
\begin{aligned}
R_T &= \mathrm{TOA}_{\mathrm{net}} + \mathrm{OLR} \\
F_S &= R_{\mathrm{short}} + R_{\mathrm{long}} + H_s + H_l
\end{aligned}
\end{equation}

\noindent
Where $\mathrm{TOA}_{\mathrm{net}}$ is the top-of-atmosphere net solar radiation, $\mathrm{OLR}$ is outgoing longwave radiation. $R_{\mathrm{short}}$, $R_{\mathrm{long}}$, $H_s$, and $H_l$ are the surface net solar radiation, surface net longwave radiation, surface net sensible heat flux, and surface net latent heat flux, respectively. Frictional heating is ignored in $F_S$.

For global sum, the divergence term in equation (\ref{sec2_eq12}) is zero, and the global sum of the tendency of $A$ is balanced by its energy sources and sinks, subject to a residual term:

\begin{equation}\label{sec2_eq14}
\overline{R_T} - \overline{F_S} -\overline{\left(\frac{\partial A}{\partial t}\right)} = \epsilon_A
\end{equation}

\noindent
Here, the net energy flux is computed as $\overline{R_T} - \overline{F_S}$ because both terms have downward as positive. The downward on the top of the atmosphere means the energy goes ``into'' the atmosphere, but the downward on the surface mean the energy ``leaves'' the atmosphere. This is different from equation \ref{sec2_eq9} where both sources and sinks are at the surface.

The air temperature ($T$) can be corrected to ensure thermal energy ($C_pT$) closes the energy budget, forcing $\epsilon_A=0$:

\begin{equation}\label{sec2_eq15}
\begin{aligned}
\overline{A^*\left(t_1\right)} = \overline{A\left(t_0\right)} + {\Delta t}\left(\overline{R_T} - \overline{F_S}\right), \quad \gamma = \frac{\overline{A^*\left(t_1\right)}}{\overline{A\left(t_1\right)}} \\
T^*\left(t_1\right) = \gamma T\left(t_1\right) + \frac{\gamma-1}{C_p}\left[L_v q\left(t_1\right)+\Phi_s+k\left(t_1\right)\right]
\end{aligned}
\end{equation}

\noindent
Where $\overline{A^*\left(t_1\right)}$ is the corrected global sum of total atmospheric energy, $\gamma$ is the multiplicative correction ratio. The same $\gamma$ is applied to $T$ at all grid cells and pressure levels.

\subsection{Physical constants, limitations, and the impact of numerical precision}\label{A15}

\begin{table}\label{tab2}
\begin{center}
\caption{The values of physical constants in this study.}
\renewcommand{\arraystretch}{1.2}
\begin{tabularx}{\textwidth}
{>{\centering\arraybackslash}X c c c}
\specialrule{1.5pt}{0pt}{3pt}
Name & Abbreviation & Value & Units \\ 
\midrule
Radius of Earth & $R$ & 6371000 & $\mathrm{m}$ \\
Gravity of Earth & $g$ & 9.80665 & $\mathrm{m \cdot s^{-2}}$ \\
Density of Water & $\rho$ & 1000.0 & $\mathrm{kg \cdot m^{-3}}$ \\
Latent Heat of Vaporization\textsuperscript{a} & $L_v$ & $\mathrm{2.501 \times 10^6}$ & $\mathrm{J \cdot kg^{-1}}$ \\
Heat Capacity of Constant Pressure for Dry Air & $C_{pd}$ & 1004.64 & $\mathrm{J \cdot kg^{-1} \cdot K^{-1}}$ \\
Heat Capacity of Constant Pressure for Water Vapor & $C_{pv}$ & 1810.0 & $\mathrm{J \cdot kg^{-1} \cdot K^{-1}}$ \\
\specialrule{1.5pt}{3pt}{0pt}
\end{tabularx}
\end{center}
\textsuperscript{a} Value obtained on 273.15 K.
\end{table}

Physical constants used in the conservation schemes above are summarized in Table \ref{tab2}. $\rho$ is used for converting the units of evaporation and total precipitation, other constants are used in the calculations directly.

A key limitation of the conservation schemes is the representation of atmospheric moisture. The specific total water ($q$) of this study is the combination of specific humidity ($q_v$), cloud liquid water content ($q_{cl}$), and rainwater content ($q_r$). For the global dry air mass and total precipitable water calculation, cloud ice ($q_{cl}$) and snow water content ($q_s$) are ignored. For the estimation of total atmospheric energy, the latent heat energy is approximated using $L_vq$. A more accurate formulation would be $\left(L_v +L_i\right) q_v + L_i \left(q_{cl}+q_r\right)$, where $L_i$ represents the latent heat of fusion. Similarly, the specific heat capacity of air at constant pressure $C_p$ is simplified as $C_{pd}(1-q)+C_{pv}q$. A corrected formulation would include $C_{pd}(1-q_v-q_{cl}-q_r-q_{ci}-q_s)+C_{pv}q_v+C_l\left(q_{cl}+q_r\right)+C_i\left(q_{ci}+q_s\right)$, where $C_l$ and $C_i$ are the heat capacities of liquid water and ice at constant pressure, respectively. In addition, $L_v$ is treated as a constant, its variation with the environmental air temperature is ignored.

Numerical precision poses another challenge to the performance of the conservation schemes. The derivation of correction ratios involves numerical integration and spatial discretization; they are less accurate under single precision (float32). As a result, residual terms cannot be closed perfectly. Switching to float64 can mitigate this issue but potentially cause slowdowns. The limitation of numerical precision also makes it difficult to model individual atmospheric moisture components, which yields the use of specific total water in this study. The mixing ratios of liquid and solid phase moisture can be extremely low at certain levels (e.g., liquid water at temperatures below 273.15 K). These low values cannot be represented effectively under a single precision.

The impact of the limitations above is minor. The values of $q_{ci}$ and $q_s$ are very small compared to $q_v$. The simplified formulations for latent heat energy and $C_p$ closely resemble their more accurate counterparts, including their temporal evolution (latent heat energy is also a small component compared to thermal energy). More than 90\% of the conservation residuals can still be closed under single precision. The above points are proved during the comparisons of conservation behaviors between the pre-processed 1.0$^\circ$ ERA5 (i.e., using the simplified formula) and the original 0.25$^\circ$ ERA5 (i.e., using the full formula) in Section \ref{sec41}. We found that, for AIWP models trained on the ERA5, the conservation schemes, although with simplified formulas, can adjust the conservation behaviors of the model effectively.  

\section{Training objective}\label{A2}

In this study, the training objective is defined as follows:

\begin{equation}\label{equ_loss}
\mathcal{L}_{\mathrm{MSE}} = \frac{1}{N}\sum_{n=0}^N \frac{1}{l}\sum_{t=0}^l \frac{1}{G}\sum_{i=0}^G \frac{1}{V}\sum_{j=0}^V \left[a(i)s(j)w(j) \left(X_{i,j}^{n, t} - Y_{i,j}^{n,t}\right)\right]
\end{equation}

\noindent
Where $n=\left\{0, 1, \cdots N\right\}$ is the index of training batches. In this study, $N=32$. $l=\left\{0, 1, \cdots L\right\}$ is the index of forecast lead time. For single-step training, $l=L=0$. For multi-step training, $l$ ranges from 0 to 11. $i=\left\{0, 1, \cdots G\right\}$ is the index of grid cells. In this study, $G=181\times 360$. $j=\left\{0, 1, \cdots V\right\}$ is the index of variables. In this study $V=77$. $a\left(i\right)$ is the latitude-based weighting. It is defined as $a\left(i\right)=\cos(\phi_i)$, $\phi_i$ is the latitude of each grid cell. 

$s\left(j\right)$ is the per-variable-level inverse variance weights. It uses the tendency of the time difference of z-scored variables. Given a z-scored variable $X_j$, its tendency is calculated as:

\begin{equation}
\Delta X_j = \left\{ X_j(t=1) - X_j(t=0), \ldots, X_j(t+1) - X_j(t) \right\}
\end{equation}

\noindent
Where $t$ is the index of time. 

Per-variable-level inverse variance weights are the inverse of the standard deviation ($\sigma$) of $\Delta X_j$:

\begin{equation}
s\left(j\right) = \frac{1}{\sigma\left(\Delta X_j\right)}
\end{equation}

\noindent
Per-variable-level inverse variance weights are higher for variables that vary more strongly over space than over time, such as geopotential height and mean sea level pressure. These variables typically converge slower during training. Implementing $s\left(j\right)$ can help resolve this problem. Following \citeA{lam2023learning}, per-variable-level inverse variance weights are applied to both prognostic and diagnostic variables.

$w\left(j\right)$ is per-variable-level loss weight. It is related to the vertical level of a given variable. In this study, $w\left(j\right)$ is defined as follows:

\begin{itemize}
    \item Upper-air variables below 300 hPa: $w=0.169$
    \item Upper-air variables on $\left\{1, 50, 150, 200, 250, 300\right\}$ hPa levels:
    \[
        w=\left\{1.69\times 10^{-4}, 0.00844, 0.0253, 0.0337, 0.0422, 0.105\right\}
    \]
    \item Prognostic single-level variables: $w=0.170$
    \item Diagnostic variables: $w=0.100$
\end{itemize}

\noindent
Here, 300 hPa is used as a separation threshold because this is roughly the bottom of the tropopause. Variables above and below the tropopause may have distinct spatiotemporal signals. The focus of AIWP models is typically quantities below the tropopause. Thus, variables below 300 hPa are up-weighted. The weights of diagnostic variables are lower than that of the prognostic single-level variables. This is because diagnostic variables are not involved in the iterative forecast directly (i.e., they are output-only variables), so their relative importance is weighed lower. 

The training objective defined in equation \ref{equ_loss} is used for both FuXi-base and FuXi-physics. For FuXi-physics, its first few training epochs may exhibit some instabilities due to the strong correction effects of the conservation schemes. This is relatively rare but can be avoided by starting the single-step training from the warm-up checkpoint of FuXi-base.

\section{Verification methods}\label{A3}

\subsection{Verifying the conservation behaviors}

This section summarizes the verified conservation quantities in Section \ref{sec41}. Explanations of the equations and terms are available in Appendix \ref{A1} as part of the derivations of the conservation schemes. The purpose of this section is to provide an emphasis on which term is verified in the manuscript. Some terms are treated differently for the pre-processed 1.0$^\circ$ ERA5 and the 0.25$^\circ$ original ERA5. This is because one of the verification goals is to examine the impact of the simplified formulas made in Appendix \ref{A1}.

For the verification of global dry air mass conservation, two quantities are computed: the global dry air mass content ($\overline{M_d}$) and the global dry air mass residual ($\epsilon_d$). For the pre-processed 1.0$^\circ$ ERA5 and the two FuXi runs, they are defined as follows:

\begin{equation}\label{sec3_eq1}
\overline{M_d} = \text{SUM}\left[\frac{1}{g}\int_{p_1}^{p_0}{\left(1-q\right)}dp\right],\quad \epsilon_d = \frac{\partial}{\partial t}\overline{M_d}
\end{equation}

\noindent
Where $\text{SUM}$ is the global weighted sum (see equation \ref{sec2_eq2}), and the pressure level integral is computed using the trapezoidal rule. $q$ is the specific total water.

For the original 0.25$^\circ$ ERA5, $\overline{M_d}$ is computed using all moisture components:

\begin{equation}\label{sec3_eq2}
\overline{M_d} = \text{SUM}\left[\frac{1}{g}\int_{p_1}^{p_0}{\left(1-q_v-q_{cl}-q_r-q_{ci}-q_s\right)}dp\right]
\end{equation}

\noindent
Where the moisture components are specific humidity ($q_v$), cloud liquid water content ($q_{cl}$), rainwater content ($q_r$), cloud ice ($q_{cl}$), and snow water content ($q_s$). Equation \ref{sec3_eq2} is more accurate but requires separated modeling of all components. As discussed in Appendix \ref{A14}, this is practically impossible. Thus, equation \ref{sec3_eq2} is applied to the original 0.25$^\circ$ ERA5 only.

In this study, $\overline{M_d}$ has the unit of kg, i.e., it is the mass of dry air for the entire atmosphere, which is expected to be a constant. The absolute value of $\epsilon_d$ is visualized, which reveals the amount of modeled dry air mass that violates the conservation.

For the verification of global moisture budget conservation, the following quantities are computed: global total precipitable water tendency, global total precipitation amount ($\overline{P}$), global evaporation amount ($\overline{E}$), and the conservation residuals ($\epsilon_m$). All with the unit of kg. For the pre-processed 1.0$^\circ$ ERA5 and the two FuXi runs, they are defined as follows:

\begin{equation}\label{sec3_eq3}
\begin{aligned}
\overline{E} &= \text{SUM}\left(E\right) \\[6pt]
\overline{P} &= \text{SUM}\left(P\right) \\[6pt]
\overline{\left(\frac{\partial M_v}{\partial t}\right)} &= \text{SUM}\left(\frac{1}{g}\frac{\partial}{\partial t}\int_{p_1}^{p_0}{q}dp\right) \\[6pt]
\epsilon_m &= -\overline{\left(\frac{\partial M_v}{\partial t}\right)} - \overline{E} - \overline{P}
\end{aligned}
\end{equation}

\noindent
Where $E$ and $P$ are evaporation and total precipitation with units of $\mathrm{kg\cdot m^{-2} \cdot s^{-1}}$. Note that the global weighted sum is applied after the tendency calculation.

Similar to equation \ref{sec3_eq2}, for the 0.25$^\circ$ original ERA5, all moisture components are applied to replace the specific total water ($q$). The absolute value of $\epsilon_m$ is visualized; it measures the mismatch between the tendency of total precipitable water in the atmosphere and the net evaporation/total precipitation.

For the verification of global total atmospheric energy conservation, the following quantities are computed: global total atmospheric energy tendency, global net energy fluxes at the top of the atmosphere and the surface, and the conservation residuals ($\epsilon_A$). All with the unit of W. The tendency term is calculated as follows:

\begin{equation}\label{sec3_eq4}
\overline{\left(\frac{\partial A}{\partial t}\right)} = \text{SUM}\left[\frac{1}{g}\frac{\partial}{\partial t}\int_{p_1}^{p_0}{\left(C_pT+L_v q+\Phi_s+k\right)}dp\right]
\end{equation}

\noindent
As explained in Appendix \ref{A15}, for the 1.0$^\circ$ pre-processed ERA5 and the two FuXi runs, The latent heat energy is simplified as $L_v q$, and $C_p$ is simplified as $C_{pd}(1-q)+C_{pv}q$. For the 0.25$^\circ$ original ERA5, their full forms are: $\left(L_v +L_i\right) q_v + L_i \left(q_{cl}+q_r\right)$ and $C_p = C_{pd}(1-q_v-q_{cl}-q_r-q_{ci}-q_s)+C_{pv}q_v+C_l\left(q_{cl}+q_r\right)+C_i\left(q_{ci}+q_s\right)$, with $L_i=333700\ \mathrm{J\cdot kg^{-1}}$ taken from 273.15 K and as a constant.

The energy fluxes and conservation residuals are computed as follows:

\begin{equation}\label{sec3_eq5}
\begin{aligned}
\overline{R_T} &= \text{SUM}\left(\mathrm{TOA}_{\mathrm{net}} + \mathrm{OLR}\right) \\[6pt] 
\overline{F_S} &= \text{SUM}\left(R_{\mathrm{short}} + R_{\mathrm{long}} + H_s + H_l\right) \\[6pt]
\epsilon_A &= \overline{R_T} - \overline{F_S} - \overline{\left(\frac{\partial A}{\partial t}\right)}
\end{aligned}
\end{equation}

\noindent
The global net energy fluxes that combine the top of the atmosphere and the surface is $\overline{R_T}-\overline{F_S}$. Note that the global weighted sum is applied after the aggregation of individual energy sources and sinks.

\subsection{Deterministic verification}

This section summarizes the deterministic verification metrics used in this study. Technical details of RMSE and zonal energy spectrum have been summarized in \citeA{schreck2024community}. The main focus of this section is SEEPS.

The implementation of SEEPS in this study is consistent with \citeA{rasp2024weatherbench}. Given precipitation forecasts and verification targets, they are converted to deterministic ``dry'', ``light'', and ``heavy'' precipitation categories. The separation between dry and light precipitation is 0.1 mm per day. The separation between light and heavy precipitation is the 66th percentile value computed from the non-dry days of each day-of-year within a climatological period. In this study, it is 1990-2019. The converted categorical forecast is verified using a 3-by-3 contingency table for each location, forecast lead time, and all initialization times. SEEPS is the matrix multiplication between the contingency table and the following score matrix:

\begin{equation}
S = \begin{pmatrix}
0 & \frac{1}{1-p} & \frac{4}{1-p} \\
\frac{1}{p} & 0 & \frac{3}{1-p} \\
\frac{1}{p}+\frac{3}{2+p} & \frac{3}{2+p} & 0
\end{pmatrix}
\end{equation}

\noindent
Where $p$ is the climatological probability of dry days. Grid cells with $p\in\left(0.1, 0.85\right)$ are included. The SEEPS described above is computed on all available grid cells, and the domain-averaged SEEPS is computed using the cosine latitude weighting (see Appendix \ref{A2}). Further details of SEEPS and its advantages are available in \citeA{rodwell2010new}.

%
%

\section*{Open Research Section}
The ERA5 reanalysis data for this study can be accessed through the NSF NCAR Research Data Archive at \url{https://rda.ucar.edu/datasets/d633000/} and the Google Research, Analysis-Ready, Cloud Optimized (ARCO) ERA5 at \url{https://cloud.google.com/storage/docs/public-datasets/era5}. The IFS-HRES forecasts of this study are obtained from Weatherbench2; they are available at \url{https://weatherbench2.readthedocs.io/en/latest/data-guide.html}. The neural networks described here and the simulation code used to train and test the models are archived at \url{https://github.com/NCAR/miles-credit}. The verification and data visualization code of this study is archived at \url{https://github.com/yingkaisha/CREDIT-physics-run}. 

\section*{Acknowledgments}
This material is based upon work supported by the NSF National Center for Atmospheric Research, which is a major facility sponsored by the U.S. National Science Foundation under Cooperative Agreement No. 1852977. This research has also been supported by NSF Grant No. RISE-2019758. We would like to acknowledge high-performance computing support from Derecho and Casper \citeA{Cheyenne} provided by the Computational and Information Systems Laboratory, NCAR, and sponsored by the National Science Foundation. YS thanks Dr. Oliver Watt-Meyer for helpful conservations.

\subsection*{Author Contributions}
YS: Conceptualization, Methodology, Software, Validation, Formal Analysis, Investigation, Data Curation, Writing—Original Draft Preparation, Writing—Review and Editing, Visualization.
JSS: Software, Data Curation, Writing—Review and Editing. 
WEC: Software, Data Curation, Writing—Review and Editing.
DJG: Conceptualization, Software, Writing—Review and Editing, Supervision, Project Administration, Funding Acquisition.

\bibliographystyle{apacite}
\bibliography{reference_arXiv}

\end{document}



\section{Data pre-processing}\label{sec1}

\subsection{Mass-conserved vertical downsampling}\label{sec11}

For a given quantity $X(p)$ that varies on a set of pressure levels $p=\left\{p_0, p_1, \ldots, p_N\right\}$, it can be down-sampled to a smaller set of pressure levels $p'=\left\{p'_0, p'_1, \ldots, p'_M\right\}$. When $p'\in p$, $p'_0 = p_0$, $p'_M=p_N$, this downsampling can be conducted to conserve the pressure level integral of $X$.

This mass-conserved downsampling produces $X'$ on $M-1$ levels with each value $X'\left(p'_{M-1}, p'_M\right)$ represents the area-averaged state between level $p_{M-1}$ and $p_M$:

\begin{equation}
\begin{aligned}
\Delta X &= \left\{\frac{X\left(p_0\right)+X\left(p_1\right)}{2}, \ldots, \frac{X\left(p_{N-1}\right)+X\left(p_{N}\right)}{2} \right\}\\
\Delta p &= \left\{\frac{p_0+p_1}{2}, \ldots, \frac{p_{N-1}+p_N}{2}\right\}\\
X'\left(p'_{M-1}, p'_M\right) &= \frac{1}{p'_M - p'_{M-1}}\sum_{i = i\left(p=p'_{M-1}\right)}^{i\left(p=p'_M\right)-1}{\left(\Delta p \Delta X\right)_{i}}
\end{aligned}
\end{equation}

\noindent
Where $i$ is the index of $p$ between $p'_{M-1}$ and $p'_M$: $p\left(i\right) \in \left[p_{M-1}, p_M\right)$. $i$ is ranged from  $i\left(p=p'_{M-1}\right)$ to $i\left(p=p'_M\right)-1$, because the length of $\Delta X$ and $\Delta p$ is $N-1$, i.e., one element short from the original pressure levels.

After down-sampling, the pressure level integral of $X'$ is computed as $X'\Delta p'$; the value is the same as the pressure level integral of $X$ on $p$ levels using the trapezoidal rule.

\section{Numerical computation of pressure level data}

\subsection{Global weighted sum}

For a given quantity $X(\phi, \lambda)$ that varies by latitude ($\phi$) and longitude ($\lambda$), its global weighted sum $\overline{X}$ is computed as follows:

\begin{equation}\label{sec2_eq1}
\overline{X} = \int_{-\pi/2}^{\pi/2} \int_{0}^{2\pi} X \cdot R^2 \cdot d\left(\sin \phi \right) d\lambda
\end{equation}

\noindent
Where $R$ is the radius of the earth. For gridded data, equation (\ref{sec2_eq1}) can be written in discrete form:

\begin{equation}\label{sec2_eq2}
\overline{X} = \sum_{i_\phi=0}^{N_\phi} \sum_{i_\lambda=0}^{N_\lambda} {\left[X \cdot R^2 \cdot \Delta\left(\sin \phi \right) \cdot \Delta \lambda\right]}_{i_\phi,i_\lambda}
\end{equation}

\noindent
Where $i_\phi=\left\{0, 1, \ldots, N_\phi\right\}$ and $i_\lambda=\left\{0, 1, \ldots, N_\lambda\right\}$ are indices of latitude and longitude, respectively. $\Delta\left(\sin \phi \right)$ and $\Delta \lambda$ are computed as grid spacings; they can be estimated using second-order difference for central grid cells and forward difference for edge grid cells.

Hereafter, the global weighted sum is denoted as $\overline{X} = \text{SUM}\left(X\right)$ 

\subsection{Pressure level integrals}

For a given quantity $X(z)$ that varies by height $z$, its mass-weighted vertical integral can be converted to pressure level integral using hydrostatic equation:

\begin{equation}\label{sec2_eq3}
\int_{0}^{\infty}{\rho X}dz = \frac{1}{g}\int_{p_s}^{0}Xdp \approx \frac{1}{g}\int_{p_0}^{p_M}Xdp
\end{equation}

\noindent
Where $p_s$ is surface pressure. The vertical integral of $X$ from the surface to $p=0$ is approximated by the range of pressures that are available from a pressure level dataset, typically 1000-1 hPa. The pressure level integral in equation (\ref{sec2_eq3}) is discretized using the trapezoidal rule. For its vertically downsampled version $X'$, as introduced in Section \ref{sec11}, the pressure level integral is further simplified to $X'\Delta p'$.

\section{Global mass and energy fixes}

\subsection{Fix the conservation of global dry air mass}

For a given air column, the tendency of its dry air mass is explained by the divergence of its vertically integrated mass flux:

\begin{equation}\label{sec2_eq4}
\frac{1}{g}\frac{\partial}{\partial t}\int_{p_0}^{p_1}{\left(1-q\right)}dp = -\mathbf{\nabla} \cdot \frac{1}{g} \int_{p_0}^{p_1}{\left[\left(1-q\right)\mathbf{v}\right]}dp
\end{equation}

For global sum, the divergence term in equation (\ref{sec2_eq4}) is zero for incompressible atmosphere. Thus, the total amount of global dry air mass ($\overline{M_d}$) is conserved regardless of time:

\begin{equation}\label{sec2_eq5}
\frac{\partial}{\partial t}\overline{M_d} = \frac{\partial}{\partial t}\text{SUM}\left(\frac{1}{g}\int_{p_0}^{p_1}{\left(1-q\right)}dp\right) = \epsilon
\end{equation}

\noindent
Where $\epsilon$ is a residual term that violates the conservation due to numerical computation.

Given two time steps $\Delta t = t_1 - t_0$ with $t_0$ representing the analyzed initial condition and $t_1$ representing an arbitrary forecasted time, equation (\ref{sec2_eq5}) can be discretized as:

\begin{equation}\label{sec2_eq6}
\overline{M_d\left(t_0\right)} - \overline{M_d\left(t_1\right)} = \epsilon
\end{equation}

\noindent
Based on the definition of $\overline{M_d}$, $q$ can be modified to force $\epsilon=0$ using a multiplicative ratio:

\begin{equation}\label{sec2_eq7}
q^*\left(t_1\right) = 1 - \left[1 - q\left(t_1\right)\right] \frac{\overline{M_d\left(t_0\right)}}{\overline{M_d\left(t_1\right)}}
\end{equation}

\noindent
Where $q^*\left(t_1\right)$ is the corrected $q$ on forecasted time $t_1$. Note that the same multiplicative correction is applied to $q$ on all grid cells and pressure levels.

\subsection{Fix global moisture budget}

For a given air column, the tendency of its total column water vapor ($M_v$) is explained by the divergence of its vertically integrated moisture flux, precipitation, and evaporation:

\begin{equation}\label{sec2_eq8}
\frac{\partial}{\partial t}M_v = \frac{1}{g}\frac{\partial}{\partial t}\int_{0}^{p_s}{q}dp = -\mathbf{\nabla} \cdot \frac{1}{g} \int_{0}^{p_s}{\left(\mathbf{v}q\right)}dp - E - P
\end{equation}

\noindent
Where $E$ and $P$ are precipitation and evaporation in flux forms with units of $\mathrm{kg\cdot m^2 \cdot s^{-1}}$. Downward flux is positive. 

For global sum, the divergence term in equation (\ref{sec2_eq8}) is zero, and the global sum of total column water vapor ($\overline{M_v}$) is balanced by its corresponding $E$ and $P$ terms, subject to a residual that violates the conservation relationships:

\begin{equation}\label{sec2_eq9}
\overline{\left(\frac{\partial M_v}{\partial t}\right)} + \overline{E} + \overline{P} = \epsilon
\end{equation}

\noindent
Flux form precipitation ($P$) can be modified to force $\epsilon=0$ using a multiplicative ratio:

\begin{equation}\label{sec2_eq10}
P^*\left(t_1\right) = P\left(t_1\right)\frac{\overline{P^*\left(t_1\right)}}{\overline{P\left(t_1\right)}}, \quad\overline{P^*\left(t_1\right)} = -\overline{\left[\frac{M_v\left(t_1\right) - M_v\left(t_0\right)}{\Delta t}\right]} - \overline{E\left(t_1\right)}
\end{equation}

\noindent
Where $\overline{P^*\left(t_1\right)}$ is the corrected global sum of precipitation flux that can close the moisture budget. Note that the same multiplicative ratio is applied to all grid cells.

\subsection{Fix global total energy budget}

For a given air column, the pressure level integral of its total atmospheric energy ($A$) is defined as follows:

\begin{equation}\label{sec2_eq11}
A = \frac{1}{g}\int_{p_0}^{p_1}{\left(C_pT+Lq+\Phi_s+k\right)}dp
\end{equation}

\noindent
The terms on the right side of equation (\ref{sec2_eq11}) are thermal energy, latent heat energy, gravitational potential energy, and kinetic energy, respectively. $L$ is the latent heat of condensation of water, $C_p$ is the specific heat capacity of air at constant pressure, and $\Phi_s$ is geopotential at the surface. Kinetic energy ($k$) is defined as $k=0.5\left(\mathbf{v} \cdot \mathbf{v}\right)$. Note that equation (\ref{sec2_eq11}) does not consider cloud formation and the fusion of water vapor. 

The tendency of $A$ is explained by the divergence of the vertically integrated moist static energy and kinetic energy, and other energy sources and sinks:

\begin{equation}\label{sec2_eq12}
\frac{\partial}{\partial t}A = -\mathbf{\nabla}\cdot\frac{1}{g}\int_{p_0}^{p_1}{\mathbf{v}\left(h+k\right)}dp = R_T - F_S
\end{equation}

\noindent
Where $h=C_pT+Lq+\Phi$ is moist static energy. $R_T$ and $F_S$ are net energy fluxes on the top of the atmosphere and the surface:

\begin{equation}\label{sec2_eq13}
\begin{aligned}
R_T &= \mathrm{TOA}_{\mathrm{net}} + \mathrm{OLR} \\
F_S &= R_{\mathrm{short}} + R_{\mathrm{long}} + H_{\mathrm{sensible}} + H_{\mathrm{latent}}
\end{aligned}
\end{equation}

\noindent
Where $\mathrm{TOA}_{\mathrm{net}}$ is the net solar radiation at the top of the atmosphere, OLR is the outgoing long-wave radiation (i.e., the net thermal radiation at the top of the atmosphere). $R_{\mathrm{short}}$ and $R_{\mathrm{long}}$ are the net solar radiation and the net thermal radiation at the surface, respectively. $H_{\mathrm{sensible}}$ and $H_{\mathrm{latent}}$ are sensible and latent heat fluxes at the surface, respectively. Frictional heating is ignored in $F_S$. Downward energy transport is positive.

For global sum, the divergence term in equation (\ref{sec2_eq12}) is zero, and the global sum of the time tendency of $A$ is balanced by its energy sources and sinks, subject to a residual term:

\begin{equation}\label{sec2_eq14}
\overline{\left(\frac{\partial A}{\partial t}\right)} -\overline{R_T} + \overline{F_S} = \epsilon
\end{equation}

\noindent
Air temperature ($T$) can be corrected, the resulting thermal energy can close the energy budget in equation (\ref{sec2_eq14}) and force $\epsilon=0$:

\begin{equation}\label{sec2_eq15}
\begin{aligned}
\overline{A^*\left(t_1\right)} = \overline{A\left(t_0\right)} + {\Delta t}\left(\overline{R_T} - \overline{F_S}\right), \quad \gamma = \frac{\overline{A^*\left(t_1\right)}}{\overline{A\left(t_1\right)}} \\
T^*\left(t_1\right) = \gamma T\left(t_1\right) + \frac{\gamma-1}{C_p}\left[Lq\left(t_1\right)+\Phi_s+k\left(t_1\right)\right]
\end{aligned}
\end{equation}



\noindent
Where $\overline{A^*\left(t_1\right)}$ is the corrected global sum of total atmospheric energy that can close the energy budget, $\gamma$ is the multiplicative correction ratio. Note that the same $\gamma$ is applied to $T$ on all grid cells and pressure levels.